\documentclass[12pt]{article}

\usepackage{amsmath}
\usepackage{amsfonts}
\usepackage{amssymb}

\usepackage{mathstyle}
\usepackage{flexisym}
\usepackage{breqn}

\usepackage{srcltx}

\usepackage{pgf}
\input xy
\xyoption{all}

\usepackage{hyperref}

\newtheorem{theo}{Theorem }[section]

\newtheorem{defi}{Definition}[section]

\newtheorem{rema}{Remark}[section]

\def\ad{\mathrm{ad\,}}

\def\Id{\mathrm{Id}}

\def\g{\mathfrak{g}}

\def\s{\mathfrak{s}}

\def\K{\mathbb{K}}

\def\R{\mathbb{R}}

\def\P{\mathbb{P}} \def\C{\mathbb{C}}

\def\Tr{\mathrm{Tr}}

\def\al{\alpha}

\def\de{\delta}
\def\d{\partial}
\def\ka{\kappa}
\def\la{\lambda}

\def\om{\omega}

\def\qed{$\square$}
\title{On   compatible linear and quadratic Poisson brackets on $gl(N)$}
\author{ A.~ Panasyuk$^{1}$,  T.~ Skrypnyk$^{2}$\\
{\small $^1$  Cardinal Stefan Wyszy\'{n}ski University, Warsaw, Poland} \\
 {\small  $^2$  Bogolyubov Institute for Theoretical Physics of NASU,
  Kyiv, Ukraine }}
\date{ }

\begin{document}
\bibliographystyle{plain}
\date{}
\maketitle

\begin{abstract}
In the present paper, using two constant tensors 
  on $sl(N)\otimes sl(N)$ satisfying certain linear-quadratic equation and a technique of Poisson bivectors and Schouten brackets,   we  explicitly construct quadratic Poisson bracket on the space $ (sl(N)+\mathbb{C})^*$ which is compatible with the standard Lie-Poisson brackets on $gl(N)^* \simeq (sl(N)\oplus \mathbb{C})^*$. The case of $N=3$ is considered in  details.  The relation of the proposed brackets with  generalized classical Sklyanin  algebras is  explained.

\end{abstract}

Keywords:  classical integrable system, Poisson pencil, quadratic algebra.


\section{Introduction}
Since its invention in 1978 \cite{m}  bihamiltonian structures proved their great efficiency in the theory of  classical integrable systems. Recall that a bihamiltonian structure is a pair $(\pi_1,\pi_2)$ of compatible Poisson structures, where the compatibility means that their sum also is a Poisson structure. The first integrals of an integrable system appear  as  Casimir functions of the elements of the pencil $\{\pi_1+\la \pi_2\}$.

One of the simplest and best studied   situations  \cite{bols'} is the case when the considered Poisson manifold is a linear  space and the Poisson structures are linear and constant correspondingly.
Next remarkable case \cite{pbi-Lie} is when both the  structures  $\pi_1$ and $\pi_2$ are  linear (Lie--Poisson) structures corresponding to some finite-dimensional Lie algebras $\mathfrak{g}$ and  $\mathfrak{g}'$.
Finally, much more complicated situation, which also can be found in the literature,  is the case  when the   first  structure  $\pi_1$ is a linear structure on $\mathfrak{g}^*$ and the second one, $\pi_2$, is quadratic. This situation is the main subject of study of the present paper in the physically most important case of $\mathfrak{g}=gl(N)$.

It is known (see e.g. \cite{RoubtsovSkr}) that quadratic Poisson bracket are always compatible with their linearization in the neighbourhood of any point of their Poisson space.  In the present  paper we consider the so-called centrally linearizable quadratic Poisson brackets \cite{pQuadrPoisson} when the linear bracket is obtained by the linearization  in a neighborhood of the central element of $\mathfrak{g}=gl(N)$. This automatically provides compatibility of a pair $(\pi_1,\pi_2)$ and imposes some restrictions on the possible form of $\pi_2$.

Nevertheless, in order to find out the final ansatz for the  tensor $\pi_2$  that  satisfies the Jacobi conditions and leads to centrally linearizable pencils  this information is insufficient. One  has  to use  insights from  the theory of classical skew-symmetric $r$-matrices with spectral parameters \cite{BD} and quadratic Sklyanin brackets \cite{sklyanin}. The last brackets are  written in terms  of Lax matrices which, generally speaking, are the elements of infinite-dimensional Poisson--Lie group.
Although the quadratic Sklyanin brackets  were an object of constant interest for many years, its finite-dimensional Poisson subspaces  have not been yet explicitly described for the general  case of  $\mathfrak{g}=gl(N)$ and arbitrary  $r$-matrix satisfying classical Yang--Baxter equation with spectral parameter. Moreover, it occurred that in order to define   quadratic structure on $gl(N)$ using Sklyanin brackets, the $r$-matrix  should satisfy not only the classical Yang--Baxter equation but also the so-called ``projected'' classical associative Yang--Baxter equation \cite{Polishch} (see also \cite{burbanGal} and references therein). Using this equation, quadratic Sklyanin brackets (see formula (\ref{sklbr})) and certain ansatz for the Lax matrix (see formula (\ref{LaxAnsatz})), we have derived an explicit form of centrally-linearizable quadratic  Poisson brackets on $gl(N)$ (see formula (\ref{qpb})). It is written it terms of two constant tensors $b, c\in sl(N)\otimes sl(N)$ coming from  an $sl(N)\otimes sl(N)$-valued classical $r$-matrix $r(u-v)$.  Since $r(u-v)$  satisfies the projected associative Yang--Baxter equation the tensors $b$ and $c$ satisfy certain linear-quadratic equation (see eq. (\ref{fp4})). We consider the obtained quadratic brackets to be  ultimate generalizations of the famous quadratic Poisson algebra of Sklyanin \cite{sklyanin} on $gl(2)$. Our generalization include, in particular, the  Khesin--Levin--Olshanetsky \cite{khesinLO}  elliptic $gl(N)$-generalizations of $gl(2)$ Sklyanin Poisson algebra and $gl(3)$  Poisson brackets of Sokolov \cite{SokolovEllCM}.

Since in the present paper we are mainly interested in the Poisson geometry and bihamiltonian structures, we do not concentrate on the Lax/$r$-matrix approaches and consider the obtained formula (\ref{qpb})  of the  quadratic structure on $gl(N)$ in completely autonomous way. In particular, using the standard properties of the Schouten bracket and direct and tedious  calculations we show that the Poisson bracket defined by the formula  (\ref{qpb}) is a Poisson bracket indeed, i.e  satisfies Jacobi condition, if the corresponding $b$-$c$ tensors satisfy equation (\ref{fp4}).  Although our proof is  valid for $N>2$, we assume that our result holds true also in the $gl(2)$ case. As an example, we present a generic family of solutions of equations  (\ref{fp4}) in the $N=3$ case.   An  open question remains whether all the   solutions of   equation (\ref{fp4}) come from the classical $r$-matrices, or, vice versa, do any solution of equation (\ref{fp4}) can be extended to a classical $r$-matrix (the existence question). Another related question is  whether the classical $r$-matrix is completely determined by its $b$-$c$ tensors (the uniqueness question).

It is worth to mention that there exists another approach to construction of the Lax matrices for the generalized  Sklyanin algebra \cite{ZotovKrasnov}. It is based on smart idea to use {\it quantum} associative Yang-Baxter equation  of  \cite{Polishch} for  constructing   the {\it classical} Lax matrix for  quadratic Sklyanin brackets on $gl(n)$.  All the answers   obtained in this approach  are naturally written in terms of the corresponding { quantum} $R$-matrix $R(u,v)$.  It would be interesting to compare the results obtained using  approach of \cite{ZotovKrasnov}  with the one proposed in the present paper.

The structure of the paper is as follows. Section 2 is devoted to preliminaries on the Schouten bracket, in Section 3 we formulate our Main Theorem and present some auxiliary results.  In Section \ref{srm}  we explain the relation of our construction with the generalized Sklyanin approach. Detailed analysis of the $N=3$ case  is performed in Section \ref{ssl3}. Finally, in Appendix we prove the Main Theorem.

At last we would like to remark that all our results with the exception of those from Section \ref{ssl3}, although formulated in complex analytic category, in fact are valid also over the ground field $\R$ and in real analytic category. The results of Section \ref{ssl3} rely on classification of complex representations of $sl(3,\C)$ and of complex cubic curves in $\C\P^2$.





\section{ Remarks on the Schouten bracket}
\label{Sho}

We use the standard properties of the Schouten bracket $[,]$, see \cite{marle2}, and we adopt conventions of that paper. According to  them any smooth $p$-vector  $\pi$ on a smooth manifold $M$ generates a $p$-linear operation on the space $\C^\infty(M)$ of smooth functions on $M$ by (cf. \cite[form. (20)]{marle2})
$$
(f_1,\ldots,f_p)_\pi:=\langle df_1\wedge\cdots\wedge df_p,P\rangle=(-1)^{[p/2]}\langle df_p\wedge\cdots\wedge df_1,P\rangle=(-1)^{[p/2]}[\ldots[P,f_1],\ldots,f_p].
$$
In particular, if in local coordinates  $\pi=\frac1{2}\pi_{ij}(x_{1},\ldots,x_{n})\frac{\d}{\d x_{i}}\wedge\frac{\d}{\d x_{j}}$, $\pi_{ij}=-\pi_{ji}$, then the corresponding bracket on functions, which will be standardly denoted $\{f_1,f_2\}$ is given by
$$
\{f_1,f_2\}=\pi_{ij}(x_{1},\ldots,x_{n})\frac{\d f_1}{\d x_{i}}\frac{\d f_2}{\d x_{j}}.
$$
A bivector $\pi$ is \emph{Poisson}, if $[\pi,\pi]=0$, or equivalently, the bracket $\{\cdot,\cdot\}$ satisfies the Jacobi identity, as the theorem below shows. Two Poisson bivectors  $\pi_1$ and $\pi_2$ are \emph{compatible} if their sum $\pi_1+\pi_2$ is again a Poisson bivector, or equivalently, $[\pi_1,\pi_2]=0$. Given a Poisson bivector  $\pi=\frac1{2}\pi_{ij}(x_{1},\ldots,x_{n})\frac{\d }{\d x_{i}}\wedge\frac{\d }{\d x_{j}}$ and a smooth function $f$, the vector field $\pi(f):=[\pi,f]=-\pi_{ij}(x_{1},\ldots,x_{n})\frac{\d f}{\d x_{i}}\frac{\d }{\d x_{j}}$ is called a \emph{hamiltonian vector field} with the \emph{hamiltonian} $f$.

The following result is well-known, but just to fix proportionality constants we shall prove it.
\begin{theo}\label{th}
Let $P$ be a bivector. Then
\begin{equation}\label{ghj}
(f_1,f_2,f_3)_{[P,P]}=-2\sum_{c.p.}\{\{f_1,f_2\},f_3\},
\end{equation}
where $\{f_1,f_2\}=(f_1,f_2)_P=-P(f_1)f_2$ and the sum is taken over the cyclic permutations of $f_i$.
\end{theo}

\noindent \textsc{Proof} We use the graded antisymmetry \cite[form. (10)]{marle2} and the graded Jacobi identity  \cite[form. (15)]{marle2} of the Schouten bracket. We have
\begin{align*}
& [P,P(f)]=\ad_P[P,f]=[\ad_PP,f]-[P,\ad_Pf]=[[P,P],f]-[P,P(f)],
\end{align*}
whence $[P,P(f)]=\frac1{2}[[P,P],f]$. Thus
\begin{align*}
[P,P(f_1)f_2]=\ad_P[P(f_1),f_2]=[[P,P(f_1)],f_2]+[P(f_1),P(f_2)]=\frac1{2}[[[P,P],f_1],f_2]+[P(f_1),P(f_2)]
\end{align*}
and
\begin{align*}
& \{\{f_1,f_2\},f_3\}+\{\{f_3,f_1\},f_2\}+\{\{f_2,f_3\},f_1\}=[[P,P(f_1)f_2],f_3]-\{f_2,\{f_3,f_1\}\}-\{f_1,\{f_2,f_3\}\}=\\
& \frac1{2}[[[[P,P],f_1],f_2],f_3]+[P(f_1),P(f_2)]f_3-P(f_2)P(f_3)f_1-P(f_1)P(f_2)f_3=-\frac1{2}(f_1,f_2,f_3)_{[P,P]}.
\end{align*}
\qed

\section{Formulation of the main theorem and auxiliary results}
Let $\g=gl(N)$ be the associative algebra of $N\times N$-matrices with
complex entries equipped with a nondegnerate symmetric bilinear form
$\langle X|Y\rangle=\Tr(XY)$. Let $\s=sl(N)$  be
the subspace of traceless matrices. Both the spaces form Lie algebras
with respect to the commutator $[,]$. Let $E_{ij}$, where $i,j \in 1,...,N$, be the natural basis of $\g=gl(N)$, i.e. $(E_{ij})_{ab}=\delta_{ia}\delta_{jb}$,  $E_{ij} E_{kl}=\delta_{kj}  E_{il}$. 

 The map $\phi: gl(N)\to gl(N)^{*}$,
$\phi(X)(\cdot)=\langle X|\cdot\rangle$, $X\in gl(N)$, isomorphically
maps $gl(N)$ to $gl(N)^{*}$ and $sl(N)$ to $sl(N)^{*}$.  Slightly abusing notations  we will hereafter  identify the spaces  $gl(N)$ and $gl(N)^{*}$, $sl(N)$ and  $sl(N)^{*}$.  The spaces  $gl(N)^{*}$ and $sl(N)^{*}$ (or rather their one-dimensional extensions) will be our linear Poisson manifolds $M$, where the wanted Poisson  brackets are  defined. The generic element of the space $gl(N)^{*}$ has the form $S=\sum\limits_{i,j=1}^N S_{ij} E_{ji}$, where $S_{ij}$ are the coordinate functions on  $gl(N)^{*}$. The  generic element of the space $sl(N)^{*}$ has the same form with the additional condition   $\sum\limits_{i=1}^N S_{ii}=0$.

In what follows we  will  assume that  $N>2$. The following theorem is the main result of this paper.

\begin{theo} \label{mainTH}
 Assume that elements $c,b\in  sl(N)\otimes sl(N)$, where $c=\sum\limits_{i,j,k,l=1}^N c_{ijkl}E_{ij}\otimes E_{kl}$, $c_{ijkl}=-c_{klij}$, and $b=\sum\limits_{i,j,k,l=1}^N b_{ijkl}E_{ij}\otimes E_{kl}$, $b_{ijkl}=b_{klij}$, satisfy the following equation:  

\begin{eqnarray}\label{fp4}
\sum\limits_{r=1}^N (c_{klir}c_{rjmn}+c_{ijmr}c_{rnkl}+c_{mnkr}c_{rlij})-\frac1{{N}} \sum\limits_{r,s=1}^N (\de_{ij}c_{klrs}c_{srmn}+
\de_{mn}c_{ijrs}c_{srkl}+\de_{kl}c_{mnrs}c_{srij})=\nonumber\\
(\delta_{kn}b_{mlij}+\delta_{mj}b_{inkl}+\delta_{il}b_{kjmn})-
\frac1{{N}}(\de_{mn}b_{ijkl}+\de_{ij}b_{klmn}+\de_{kl}b_{mnij})
-\frac1{{N}}(\de_{mn}b_{klij} +\de_{ij}b_{mnkl}+\de_{kl}b_{ijmn}).
\end{eqnarray}

Let
$$
\pi:= \frac1{2}\sum\limits_{i,j, k,l=1}^N\Bigl( \sum\limits_{m,n=1}^N (c_{kjnm}S_{mn}S_{il}-
c_{ilnm}S_{mn}S_{kj})+
\sum\limits_{s,t=1}^N(c_{iskt}S_{sj}S_{tl}-c_{sjtl}S_{is}S_{kt})\Bigr)\frac{\d }{\d S_{ij}}\wedge \frac{\d }{\d S_{kl}}
$$
be a quadratic bivector  on $gl^*(N)$,
$$
\pi_1:=\frac1{2}\sum\limits_{i,j, k,l =1}^N (\de_{jk}S_{il}-\de_{li}S_{kj})\frac{\d }{\d S_{ij}}\wedge \frac{\d }{\d S_{kl}}
$$
be the canonical linear (Lie--Poisson) bivector on $gl^*(N)$ and
\begin{align*}
\pi_1(H)=\sum\limits_{ k,l = 1}^N\left(\sum\limits_{s, m,n, n'm' = 1}^N (-c_{ksnm}c_{mnn'm'}S_{m'n'}S_{sl}+
c_{slnm}c_{mnn'm'}S_{m'n'}S_{ks})+\right.\\
\left.\sum\limits_{s, m,n = 1}^N2(b_{ksnm}S_{mn}S_{sl}-b_{slnm}S_{mn}S_{ks})\right)\frac{\d}{\d S_{kl }}
\end{align*}
be the hamiltonian vector field with respect to $\pi_1$ with the quadratic hamiltonian function
$$
H=
\frac1{2}\left(\sum\limits_{k,l,m,n=1}^N  2b_{kl nm}S_{mn}S_{lk}-\sum\limits_{k,l,m,n,m',n'=1}^N  c_{kl nm}c_{mnn'm'} S_{lk} S_{m'n'}\right).
$$
Then the following bivector on the space $(gl(N)\oplus \mathbb{C})^*$
\begin{equation}
\pi_{2}=\pi-\pi_{1}(H)\wedge\frac{\d}{\d S_{0}}+S_{0}\pi_{1}\label{form},
\end{equation}
 where $S_0$ is an auxiliary variable, is a Poisson bivector when restricted on the subspace $(sl(N)\oplus \mathbb{C})^*$.
\end{theo}

(See Appendix \ref{ApB} for the proof of the theorem.)

\medskip

\begin{rema}\label{rem1}\rm

Note that the  Poisson bivectors $\pi_1$ and $\pi_2$  defined in the Theorem  \ref{mainTH} yield -- upon  the trivial extension of the bracket $\{\ ,\ \}_1$ onto the variable $S_0$,   the following explicit form of the  compatible linear and quadratic  Poisson brackets on $(sl(N)\oplus \mathbb{C})^*$:
\begin{subequations}\label{lpb}
\begin{equation}
\{S_{ij},S_{kl}\}_1=(\de_{jk}S_{il}-\de_{li}S_{kj}),
\end{equation}
\begin{equation}
\{S_{0},S_{kl}\}_1=0,
\end{equation}
\end{subequations}
\begin{subequations}\label{qpb}
\begin{equation}
\{S_{ij},S_{kl}\}_2=S_0(\de_{jk}S_{il}-\de_{li}S_{kj})+ \sum\limits_{m,n=1}^N (c_{kjnm}S_{mn}S_{il}-
c_{ilnm}S_{mn}S_{kj})+
\sum\limits_{s,t=1}^N(c_{iskt}S_{sj}S_{tl}-c_{sjtl}S_{is}S_{kt}),
\end{equation}
\begin{equation}
\{S_{0},S_{kl}\}_2= \sum\limits_{s, m,n, n'm' = 1}^N (
c_{slnm}c_{mnn'm'}S_{m'n'}S_{ks} -c_{ksnm}c_{mnn'm'}S_{m'n'}S_{sl})+\sum\limits_{s, m,n = 1}^N
2(b_{ksnm}S_{mn}S_{sl}-b_{slnm}S_{mn}S_{ks}),
\end{equation}
\end{subequations}
\end{rema}
where in the formulae (\ref{lpb}) and (\ref{qpb}) the $sl(n)$-restriction, i.e. the condition  $\sum\limits_{i=1}^N S_{ii}=0$ is assumed.

\begin{rema}\label{rem1'}\rm  Observe  that the constraint    $\sum\limits_{i=1}^N S_{ii}=0$ is agreed with the brackets  (\ref{lpb}) and (\ref{qpb}). Indeed,  it is easy to show that $\{S_{ij},\sum\limits_{k=1}^N S_{kk}\}_1=0$, $\{S_{0},\sum\limits_{k=1}^N S_{kk}\}_1=0$ and
$\{S_{ij},\sum\limits_{k=1}^N S_{kk}\}_2=0$, $\{S_{0},\sum\limits_{k=1}^N S_{kk}\}_2=0$, $\forall i,j \in\overline{1,N}$. The proof of this fact for the bracket $\{\ ,\ \}_1$ is well-known.
The proof for the bracket  $\{\ ,\ \}_2$  is achieved  by putting $l=k$ in (\ref{qpb}), summation over $k$, making use of the skew-symmetry of the tensor $c$ and appropriate renaming of  summation indexes.
\end{rema}

\begin{rema}\label{rem1}\rm
Observe that from the explicit form of the Poisson brackets $\{\ ,\ \}_1$  and $\{\ ,\ \}_2$ it follows, in particular, that the function $S_0$ produces with respect to $\{\ ,\ \}_1$ the same hamiltonian flow as the generalized top hamiltonian $H$ with respect to the Lie--Poisson brackets  $\{\ ,\ \}_1$ since
$$
\{S_{0},S_{kl}\}_2=-\{H, S_{kl}\}_1.
$$
In other words, from the point of view of the hamiltonian dynamics, the one-dimensional extension of the phase space $sl^*(N)$ by the variable $S_0$  is equivalent ---  in the  ``optics'' of quadratic bracket --- to adding  of the  Hamiltonian defining the needed time flow
to the algebra of the initial dynamical variables.
\end{rema}

\begin{theo}\label{th0} (cf. \cite{pQuadrPoisson})
Let $M$ be a manifold, $\pi$ and $\pi_1$  two bivectors on $M$ and $H$ a smooth function om $M$. Assume $\pi_1$ is a Poisson bivector.
Then the bivector
is of the form
\begin{equation}
\pi_{2}=\pi-\pi_{1}(H)\wedge\frac{\d}{\d S_{0}}+S_{0}\pi_{1}\label{form}
\end{equation}
where $S_0$ is a coordinate on $\K$ ($\K$ is equal to $\R$ or $\C$ depending on the category), is a Poisson bivector on $M\times\K$ if and only if the following identities are satisfied:
\begin{equation}
[\pi,\pi]=2\pi_{1}(H)\wedge\pi_{1},\label{eqbasic}
\end{equation}
\begin{equation}
[\pi,\pi_{1}(H)]=0,\label{eqbasic2}
\end{equation}
and
\begin{equation}
[\pi,\pi_{1}]=0.\label{eqbasic3}
\end{equation}
Here $[,]$ is the Schouten bracket, $\pi_{1}(h)=[\pi_{1},H]$ stands for the hamiltonian vector field
corresponding to the hamiltonian function $H$.
\end{theo}

\noindent \textsc{Proof}  Since the vector fields $\pi_{1}(H)$
and $\frac{\d}{\d S_{0}}$ commute, we have $[\pi_{1}(H)\wedge\frac{\d}{\d S_{0}},\pi_{1}(H)\wedge\frac{\d}{\d S_{0}}]=0$.
Also $[S_{0}\pi_{1},S_{0}\pi_{1}]=0$, $[\pi,\frac{\d}{\d S_{0}}]=0$,
$[\pi_{1},\pi_{1}(H)]=0$, whence
\begin{align}
[\pi_{2},\pi_{2}]=[\pi,\pi]-2[\pi,\pi_{1}(H)\wedge\frac{\d}{\d S_{0}}]+2S_{0}[\pi,\pi_{1}]-2[\pi_{1}(H)\wedge\frac{\d}{\d S_{0}},S_{0}\pi_{1}]=\nonumber \\
[\pi,\pi]-2[\pi,\pi_{1}(H)]\wedge\frac{\d}{\d S_{0}}-2\pi_{1}(q)\wedge\pi_{1}+2S_{0}[\pi,\pi_{1}].\label{ff}
\end{align}
The second term is the only containing $\frac{\d}{\d S_{0}}$ and the last term is the only containing $S_0$, hence
$[\pi_{2},\pi_{2}]=0$ if and only if
(\ref{eqbasic})--(\ref{eqbasic3}) hold.
\qed

\begin{theo}\label{th2}  \cite{pQuadrPoisson} Retaining the assumptions of Theorem \ref{th0} assume additionally, that the sum of generic symplectic leaves of the bivector $\pi_1$ is dense in $M$. If dimension of the generic symplectic leaf
of the the bivector $\pi_1$ is greater or equal 6, then identity (\ref{eqbasic2})
follows from (\ref{eqbasic}) and (\ref{eqbasic3}). \end{theo}

\smallskip{}

\noindent \textsc{Proof} The graded Jacobi identity for the Schouten
bracket applied to bivectors $a,b,c$ looks standardly:
\[
[a,[b,c]]+[c,[a,b]]+[b,[c,a]]=0.
\]
Putting $a=b=c=\pi$ we get $[\pi,[\pi,\pi]]=0$. Applying the operator
$[\pi,\cdot]$ to identity (\ref{eqbasic}) and using the Leibnitz
identity for the Schouten bracket and the fact that $[\pi,\pi_{1}]=0$
we obtain
\[
[\pi,\pi_{1}(H)]\wedge\pi_{1}=0.
\]
Now fix a point $x$ in a generic symplectic leaf of $\pi_{1}$ and
use the Darboux basis for $\pi_{1}|_{x}$:
\[
\pi_{1}|_{x}=e_{1}\wedge e_{2}+e_{3}\wedge e_{4}+\cdots+e_{2n-1}\wedge e_{2n}.
\]
It is easy to see that under the restriction $2n\ge 6$ the equation
$a\wedge\pi_{1}(x)=0$, where $a$ is an unknown bivector, has only
the trivial solution. Indeed, let $e_{1},\ldots,e_{2n},e_{2n+1},\ldots,e_{m}$
be a basis of $T_{x}M$ and let $a=\sum_{i<j}a_{ij}e_{i}\wedge e_{j}$.
Then the equality $a\wedge\pi_{1}(x)=0$ is equivalent to the system
of linear equations
\begin{align*}
a_{12}+a_{34}=0, & a_{12}+a_{56}=0,\ldots, & a_{12}+a_{2n-1,2n}=0,\\
 & a_{34}+a_{56}=0,\ldots, & a_{34}+a_{2n-1,2n}=0,\\
 &  & \vdots\\
 &  & a_{2n-3,2n-2}+a_{2n-1,2n}=0,\\
 &  & a_{ij}=0,
\end{align*}
where $(i,j)$ runs through all the pairs of indices not appearing
in the equations above. This system has only the trivial solution.
\qed

\medskip

\begin{rema}\label{rem1}\rm
Observe that it is a restriction on the dimension of the generic symplectic leaf
of the the bivector $\pi_1$  in the theorem above that makes our proof valid for the case $N\geq 3$.
\end{rema}

\begin{rema}\label{remao}\rm Below we shall use the following obvious generalization of the implication guaranteed by Theorems \ref{th0} and \ref{th2}: the equalities
\begin{align}\label{eqbasic'k}&[\pi,\pi_1]=0,\\\label{eqbasic'}
&[\pi,\pi]=2\pi_1\wedge\pi_1(\ka^2 H)
\end{align}
with $\ka\not=0$ imply
$[\pi_3,\pi_3]=0$,
where
 \begin{align}\label{form1}
&\pi_3=\pi-\ka\pi_{1}(H)\wedge\frac{\d}{\d S_{0}}+\ka S_{0}\pi_{1}.
\end{align}
\end{rema}

\section{Relation with the generalized Sklyanin algebras}
\label{srm}

  Now we will briefly explain the relation of our construction with the (generalized) classical Sklyanin  algebra.  For this purpose we consider a classical $r$-matrix \cite{sklyanin,BD}, i.e $sl(N)\otimes sl(N)$-valued tensor depending on a  difference of two complex parameters,   written in the component form as follows:
   $$r(u-v)=\sum\limits_{i,j,k,l} r_{ij kl}(u-v) E_{ij}\otimes E_{kl},$$
   where the condition $\mathrm{tr}_1( r(u-v))=   \mathrm{tr}_2( r(u-v))=0$ is naturally implied.

  Let $r(u-v)$  satisfy  the skew-symmetry condition $$r^{12}(u-v)=-r^{21}(v-u),$$
  i.e.  $r_{ij kl}(u-v)=- r_{kl ij}(v-u)$, $i,j,k,l\in 1,....,N$,
  and the projected associative Yang--Baxter equation \cite{burbanGal}:
  \begin{subequations}\label{AYBE}
  \begin{equation}
  p_N\otimes p_N\otimes p_N \Bigl( r^{12}(u-v) r^{13}(u-w) + r^{13}(u-w) r^{23}(v-w)- r^{23}(v-w) r^{12}(u-v)\Bigr)=0,
  \end{equation}
  \begin{equation}
  p_N\otimes p_N\otimes p_N \Bigl( r^{12}(u-v) r^{23}(v-w) - r^{13}(u-w) r^{12}(u-v)- r^{23}(v-w) r^{13}(u-w) \Bigr)=0,
  \end{equation}
  \end{subequations}
where $p_N: gl(N)\rightarrow sl(N)$ is the orthogonal projection operator  given by
\begin{equation*}
p_N(E_{ij})=E_{ij}-\frac{1}{N}\delta_{ij}\sum\limits_{s=1}^N E_{ss}
\end{equation*}
and  $r^{12}(u-v)=\sum\limits_{i,j,k,l} r_{ij kl}(u-v) E_{ij}\otimes E_{kl} \otimes 1$, $r^{13}(u-w)=\sum\limits_{i,j,k,l} r_{ij kl}(u-w) E_{ij}\otimes 1\otimes  E_{kl} $, etc.

From equations (\ref{AYBE}) one easily derives the usual classical Yang--Baxter equations \cite{sklyanin, BD}:
 \begin{equation}
 [ r^{12}(u-v), r^{13}(u-w)] + [r^{13}(u-w), r^{23}(v-w)]+ [ r^{12}(u-v), r^{23}(v-w)])=0.
  \end{equation}
The $r$-matrix $r(u-v)$ possesses the following Laurent power  decomposition \cite{BD}:
$$
r(u-v)=\frac{\Omega}{u-v}+ c + (u-v) b + o(u-v)
$$
where the tensor $c$ is skew-symmetric and the tensor $b$ is symmetric:
$$c_{ij kl}=- c_{kl ij}, \quad b_{ij kl}=b_{kl ij}.$$
Here  $c=\sum\limits_{i,j,k,l} c_{ij kl} E_{ij}\otimes E_{kl}$, $b=\sum\limits_{i,j,k,l} b_{ij kl} E_{ij}\otimes E_{kl}$ and   $\Omega=  p_N\otimes p_N (\sum\limits_{i,j=1}^N E_{ij}\otimes E_{ji})$ is the tensor Casimir.

Using comparison of the main parts of the Laurent decompositions it is possible to show that,  tensors $c$ and $b$   on $sl(N)\otimes sl(N)$ satisfy linear-quadratic equation (\ref{fp4}) if $r$ satisfies projected associative AYBE (\ref{AYBE}) and, moreover,  brackets (\ref{qpb}) are obtained from the Sklyanin formula
\begin{equation}\label{sklbr}
\{L(u)\otimes 1_N, 1_N\otimes L(v)\}_2=[r(u-v), L(u)\otimes L(v)],
\end{equation}
 with the following ansatz for the Lax matrices:
\begin{equation}\label{LaxAnsatz}
L(u)=S_0 1_N+ \frac{1}{u}\sum\limits_{i,j=1}^n S_{ji} E_{ij}+ \sum\limits_{i,j=1}^n c_{ij kl} S_{ij} E_{kl} +  u \sum\limits_{i,j=1}^n b_{ij kl} S_{ij} E_{kl} +o(u), \qquad \sum\limits_{i=1}^N S_{ii}=0, \quad 1_N=\sum\limits_{i=1}^N E_{ii}.
\end{equation}
The Hamiltonian $H$ is calculated  as $H= \mathrm{res}_{u=0} \frac{1}{2 u} \mathrm{tr}(L^2(u))$ and the ``Hamiltonian'' $S_0$, $S_0=  \frac{1}{N}\mathrm{tr} (L(u))$.

In other words,  classical $r$-matrix satisfying  projected classical Yang--Baxter equation produces the $c$-$b$ tensors for brackets (\ref{qpb}). An  open question remains whether all the  solutions of equation (\ref{fp4}) come from the classical $r$-matrices and whether any skew-symmetric $r$-matrix satisfying (\ref{AYBE}) is completely determined by its constant and linear  part of the Laurent power series decomposition with respect to the spectral parameter.

As an illustrative example consider the following solution to equation (\ref{AYBE}) on $sl(3)$ \cite{AASzotov}:
\begin{align*}
r(u-v)=\frac{\Omega}{u-v}+E_{11}\wedge E_{21}+E_{21}\wedge E_{33}+E_{23}\wedge E_{31}-3 E_{32}\wedge E_{12}+\\
(u-v)(3E_{33}\odot E_{32}-3 E_{12}\odot E_{31}-3E_{11}\odot E_{32}+2E_{21}\odot E_{21})+\\
(u-v)^2(E_{33}\wedge E_{31}+E_{32}\wedge E_{21}+2E_{11}\wedge E_{31}+3E_{31} \wedge E_{22})+\\
(u-v)^3(2E_{21}\odot E_{31}-6E_{32}\odot E_{32})-3(u-v)^4E_{32}\wedge E_{31}+2(u-v)^5E_{31}\odot E_{31},
\end{align*}
where $E_{ij}\wedge E_{kl}=E_{ij}\otimes E_{kl}-E_{kl}\otimes E_{ij}$ and $E_{ij}\odot E_{kl}=E_{ij}\otimes E_{kl}+E_{kl}\otimes E_{ij}$. Then the constant part of this Laurent decomposition  together with the linear part satisfy (\ref{fp4}). According to the classification of Section \ref{ssl3} the constant part coincides with $\frac1{2}(-c_4+c_8)$, and corresponds to the nodal cubic curve (see case ($\mathbf{c_1}$)).

The following definition is an infinitesimal analogue of the gauge equivalence of $r$-matrices with spectral parameter.

\begin{defi}\label{gaugedef}\rm
We say that two pairs of tensors $(c,b)$ and $(c',b')$, $c,c',b,b'\in sl(N)\otimes sl(N)$, with skew-symmetric $c$ and $c'$, are gauge equivalent if there exists $X\in sl(N)$ such that $c'=c+(\ad_X\otimes\Id)\Omega, b'=b+\frac1{2}((\ad_X)^2\otimes \Id)\Omega +(\ad_X\otimes \Id)c$.
\end{defi}

\begin{rema}\rm
The classical result \cite{BD} says that any classical $r$-matrix $r(u,v)$ is gauge equivalent to a one with the dependence only on $u-v$, which at the infinitesimal level means that by a gauge transform one can always achieve symmetric $b$. It is possible to give  an independent proof of this fact based only on equations (\ref{fp4}). Thus, considering pairs of tensor $(c,b)$ with symmetric $b$ does not restrict the generality.
\end{rema}

\section{$sl(3)$-case}
\label{ssl3}

According to \cite[Table 5]{vinbergOniLieGrAlgGr} the space $\bigwedge^2(sl(3))$ of skew-symmetric tensors over the adjoint representation $Ad$ of $sl(3)$ decomposes to three irreducible components: $W$ and $W^*$ corresponding to $3\om_1$ and $3\om_2$, $\om_i$ being the fundamental weights, and the adjoint representation $Ad$ itself. The spaces $W$ and $W^*$ are 10-dimensional (the corresponding 10 weight spaces are all of multiplicity one) and we claim that for any representative $c$ of both $W$ and $W^*$ there exists a symmetric tensor $b$ such that the pair $(c,b)$ is a solution to equations (\ref{fp4}). The $Ad$ component corresponds to the tensors which are  gauge equivalent to the trivial one (see Definition \ref{gaugedef}) and can be neglected. The authors do not know any solution $(c,b)$ of equations (\ref{fp4}) with $c\in W+W^*$ and $c\not\in W$ or $c\not\in W^*$.

Below we give explicit formulas for nonzero coefficients of 10 tensors $c_\al\in W^*$, $c_\al=c_{\al,ijkl}E_{ij}\otimes E_{kl}$, $c_{\al,ijkl}=-c_{\al,klij}$, constituting a basis of the vector space $W^*$ corresponding to the weight decomposition as well as for the symmetric tensor $b$ such that $(\sum_\al y_\al c_\al,b)$ is a solution to equations (\ref{fp4}) for any $y_\al$. Recalling the schematic root and weight diagram of the Lie algebra $sl(3)$
$$
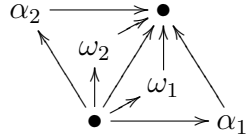
\xymatrix@C=1pt@R=1pt{
& & \al_2 \ar[rrrr] & & & & \bullet  & &\\
& & &   & \om_2 \ar[urr] &  &  &  &\\
& &  &   &  & & \om_1 \ar[uu] & &\\
 & &   & &  \bullet \ar[rruuu] \ar[rrrr] \ar[uuull] \ar[uu] \ar[urr] &  & & & \al_1 \ar[uuull]   \\
 }
$$
we enumerate the corresponding generators of  $W^*$  in the diagram
$$
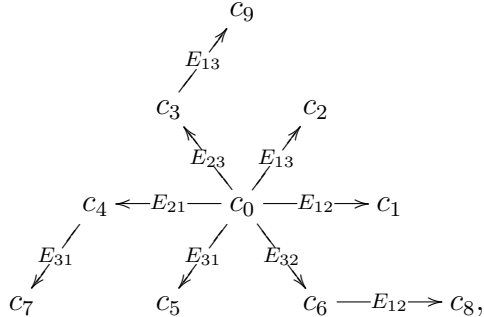
\xymatrixcolsep{1pc}\xymatrix{
 & & & c_9   &   &  &\\
& & c_3 \ar[ur]|-{E_{13}} & & c_2 & &\\
& c_4 \ar[dl]|-{E_{31}} &  &  c_0 \ar[dl]|-{E_{31}} \ar[dr]|-{E_{32}} \ar[ul]|-{E_{23}} \ar[ur]|-{E_{13}} \ar[rr]|-{E_{12}} \ar[ll]|-{E_{21}}  & & c_1 & \\
c_7 & & c_5   & & c_6 \ar[rr]|-{E_{12}} &  & c_8,}
$$
where $c_0$ is the zero weight vector and $c_9$ is the highest weight vector, vectors $c_1,\ldots,c_6$ are obtained from the $c_0$ by means of the action of the elements $E_{12},E_{13},E_{23},E_{21},E_{31},E_{32}$ respectively, and finally $c_7=E_{31}\cdot c_4,c_8=E_{12}\cdot c_6,c_9=E_{13}\cdot c_3$. Explicitly:
\begin{align*}
c_{0,1122}=c_{0,3311}=c_{0,2233}=c_{0,1221}=c_{0,3113}=c_{0,2332}=1,\\
c_{1,2212}=c_{1,1233}=c_{1,1332}=2,\\
c_{2,1333}=c_{2,2213}=c_{2,2312}=2,\\
c_{3,2311}=c_{3,2113}=c_{3,3323}=2,\\
c_{4,2111}=c_{4,3321}=c_{4,3123}=2,\\
c_{5,1131}=c_{5,3221}=c_{5,3122}=2,\\
c_{6,1132}=c_{6,1231}=c_{6,3222}=2,\\\
c_{7,2131}=6,\\
c_{8,3212}=6,\\
c_{9,1323}=6.
\end{align*}
One checks directly that the  following symmetric tensor $b=b_{ijkl}E_{ij}\otimes E_{kl}$, $b_{ijkl}=b_{klij}$, is a solution of equations (\ref{fp4}) with $c=\sum_\al y_\al c_\al$:
\begin{align*}
b_{1 1 1 1} = 2y_0^2-8y_3y_6, b_{1 2 1 1} = -4y_0y_1-4y_2y_6+12y_3y_8, b_{1 2 1 2} = 8y_1^2+24y_2y_8, \\
b_{1 3 1 1} = -4y_0y_2-4y_1y_3+12y_6y_9, b_{1 3 1 2} = 4y_1y_2-36y_8y_9, \\
b_{1 3 1 3} = 24y_1y_9+8y_2^2, b_{2 1 1 1} = 4y_0y_4+8y_3y_5, b_{2 1 1 2} = -y_0^2-4y_1y_4+4y_2y_5+4y_3y_6,\\
 b_{2 1 1 3} = -4y_2y_4-12y_5y_9, b_{2 1 2 1} = 24y_3y_7+8y_4^2, b_{2 2 1 1} = -y_0^2-4y_1y_4+4y_2y_5+4y_3y_6,\\
  b_{2 2 1 2} = 4y_0y_1+8y_2y_6, b_{2 2 1 3} = -4y_1y_3-12y_6y_9, b_{2 2 2 1} = -4y_0y_4+12y_2y_7-4y_3y_5,\\
   b_{2 2 2 2} = 2y_0^2-8y_2y_5, b_{2 3 1 1} = -4y_2y_4-12y_5y_9, b_{2 3 1 2} = -4y_1y_3-12y_6y_9, \\
   b_{2 3 1 3} = 12y_0y_9-8y_2y_3, b_{2 3 2 1} = 4y_3y_4-36y_7y_9, b_{2 3 2 2} = -4y_0y_3-4y_2y_4+12y_5y_9,\\
    b_{2 3 2 3} = 8y_3^2+24y_4y_9, b_{3 1 1 1} = 4y_0y_5+8y_4y_6, b_{3 1 1 2} = -4y_1y_5-12y_4y_8,\\
     b_{3 1 1 3} = -y_0^2+4y_1y_4-4y_2y_5+4y_3y_6, b_{3 1 2 1} = 12y_0y_7-8y_4y_5, b_{3 1 2 2} = -12y_1y_7-4y_4y_6,\\
      b_{3 1 2 3} = -12y_2y_7-4y_3y_5, b_{3 1 3 1} = 8y_5^2+24y_6y_7, b_{3 2 1 1} = -4y_1y_5-12y_4y_8, \\
       b_{3 2 1 2} = 12y_0y_8-8y_1y_6, b_{3 2 1 3} = -4y_2y_6-12y_3y_8, b_{3 2 2 1} = -12y_1y_7-4y_4y_6, \\
       b_{3 2 2 2} = 4y_0y_6+8y_1y_5, b_{3 2 2 3} = -y_0^2+4y_1y_4+4y_2y_5-4y_3y_6, b_{3 2 3 1} = 4y_5y_6-36y_7y_8,\\
        b_{3 2 3 2} = 24y_5y_8+8y_6^2, b_{3 3 1 1} = -y_0^2+4y_1y_4-4y_2y_5+4y_3y_6, b_{3 3 1 2} = -4y_2y_6-12y_3y_8,\\
         b_{3 3 1 3} = 4y_0y_2+8y_1y_3, b_{3 3 2 1} = -12y_2y_7-4y_3y_5, b_{3 3 2 2} = -y_0^2+4y_1y_4+4y_2y_5-4y_3y_6,\\
          b_{3 3 2 3} = 4y_0y_3+8y_2y_4, b_{3 3 3 1} = -4y_0y_5+12y_1y_7-4y_4y_6,\\
           b_{3 3 3 2} = -4y_0y_6-4y_1y_5+12y_4y_8, b_{3 3 3 3} = 2y_0^2-8y_1y_4.
\end{align*}

Moreover, it is easy to get a classification of the tensors $c\in W^*$ and, as a consequence, of the corresponding quadratic Poisson structures under the action of $SL(3)$. Indeed, the space $W^*$ is isomorphic to the space of cubic forms on $\C^3$, where the isomorphism is obtained by the natural identification of the basis $\{c_\al\}$ with the following basis of cubic forms, which we organize into the diagram
$$
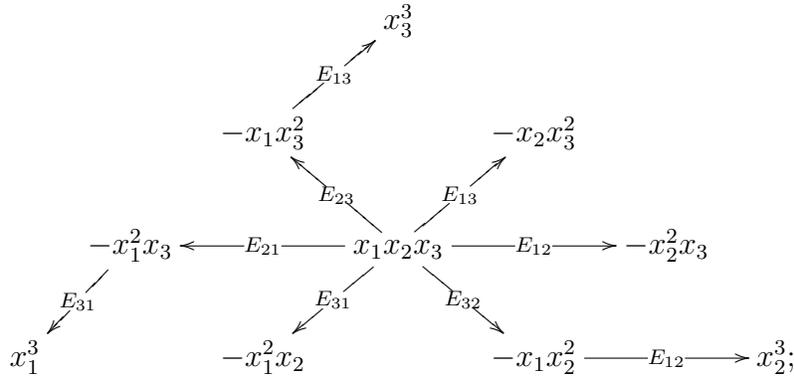
\xymatrixcolsep{1pc}\xymatrix{
 & & & x_3^3   &   &  &\\
& & -x_1 x_3^2 \ar[ur]|-{E_{13}} & & -x_2x_3^2 & &\\
& -x_1^2 x_3 \ar[dl]|-{E_{31}} &  &  x_1x_2x_3 \ar[dl]|-{E_{31}} \ar[dr]|-{E_{32}} \ar[ul]|-{E_{23}} \ar[ur]|-{E_{13}} \ar[rr]|-{E_{12}} \ar[ll]|-{E_{21}}  & & -x_2^2 x_3 & \\
x_1^3 & & -x_1^2 x_2   & & -x_1 x_2 ^2 \ar[rr]|-{E_{12}} &  & x_2^3;}
$$
here $x_i$ are the standard coordinates on $\C^3$. Recall  \cite{kraft} that the normal forms of cubic forms are as
follows: ($\mathbf{a_1}$) $x_1^3$; ($\mathbf{a_2}$) $x_1^2x_2$; ($\mathbf{a_3}$) $x_1x_2(x_1+x_2)$;
($\mathbf{a_4}$) $tx_1x_2x_3$, $t\in\C^*$; ($\mathbf{b_1}$) $(x_1^2-x_2x_3)x_2$; ($\mathbf{b_2}$)
$t(x_1^2-x_2x_3)x_1$, $t\in\C^*$; ($\mathbf{c_1}$) $x_2^2x_3-x_1^3$; ($\mathbf{c_2}$)
$t(x_2^2x_3-x_1^3-x_1^2x_3)$, $t\in\C^*$; ($\mathbf{c_3}$) $t(x_1^3+x_2^3+x_3^3)+ax_1x_2x_3$,
$t\in\C^*,a\in\C$. Case ($\mathbf{c_3}$)
corresponds to the smooth normal elliptic curve in $\C\P^2$ and  coincides with the Hesse normal
form, cases ($\mathbf{c_2}$)  and ($\mathbf{c_3}$) correspond to the nodal and cuspidal curves respectively and other cases to deeper degenerations.

Accordingly we get the following normal forms of solutions of equations (\ref{fp4}) (and the corresponding quadratic Poisson bivectors):
\begin{enumerate}\item[($\mathbf{a_1}$)] $c_{2131}=6,b=0$ ($y_7=1$); \item[($\mathbf{a_2}$)] $c_{1131}=c_{3221}=c_{3122}=-2,b_{3131}=8$  ($y_5=-1$);
\item[($\mathbf{a_3}$)] $c_{1131}=c_{3221}=c_{3122}=-2,b_{3131}=8,b_{3232}=8,b_{3231}=4$ ($y_5=-1,y_6=-1$);
    \item[($\mathbf{a_4}$)] $c_{1122}=c_{3311}=c_{2233}=c_{1221}=c_{3113}=c_{2332}=t,b_{1 1 1 1} = 2t^2,b_{2 1 1 2} = -t^2,b_{2 2 1 1} = -t^2,b_{2 2 2 2} = 2t^2,b_{3 1 1 3} = -t^2,b_{3 2 2 3} = -t^2,b_{3 3 1 1} = -t^2,b_{3 3 2 2} = -t^2,b_{3 3 3 3} = 2t^2 $ ($y_0=t$);
    \item[($\mathbf{b_1}$)] $c_{1131}=c_{3221}=c_{3122}=-2,c_{1333}=c_{2213}=c_{2312}=2,b_{1313}=8,b_{2 1 1 2} = -4,b_{2 2 1 1} = -4,b_{2 2 2 2} = 8,b_{3 1 1 3} = 4,b_{3 1 3 1} = 8,b_{3 2 2 3} = -4,b_{3 3 1 1} = 4,b_{3 3 2 2} = -4$ ($y_5=-1,y_2=1$);
        \item[($\mathbf{b_2}$)] $c_{1122}=c_{3311}=c_{2233}=c_{1221}=c_{3113}=c_{2332}=-t,c_{2131}=6t,b_{1 1 1 1} = 2t^2, b_{2 1 1 2} = -t^2, b_{2 2 1 1} = -t^2, b_{2 2 2 2} = 2t^2, b_{3 1 1 3} = -t^2,b_{3121}=-12t^2, b_{3 2 2 3} = -t^2, b_{3 3 1 1} = -t^2, b_{3 3 2 2} = -t^2,  b_{3 3 3 3} = 2t^2 $ ($y_0=-t,y_7=t$);
           \item[($\mathbf{c_1}$)] $c_{2212}=c_{1233}=c_{1332}=-2,c_{2131}=-6,b_{1212}=8,b_{3122}=-12,b_{3221}=-12,b_{3321}=12$ ($y_1=-1,y_7=-1$);
               \item[($\mathbf{c_2}$)] $c_{2212}=c_{1233}=c_{1332}=-2t,c_{2111}=c_{3321}=c_{3123}=2t,c_{2131}=-6t,b_{1 2 1 2} = 8t^2, b_{2 1 1 2} =4 t^2, b_{2 1 2 1} = 8t^2, b_{2 2 1 1} = 4t^2,  b_{3 1 1 3} = -4t^2,b_{3122}=-12t^2, b_{3 2 2 1} = -12t^2, b_{3223}=-4t^2,b_{3 3 1 1} = -4t^2, b_{3 3 1 1} = -4t^2, b_{3 3 2 2} = -4t^2, b_{3 3 3 3} = 8t^2$ ($y_1=y_7=-t,y_4=t$);
                   \item[($\mathbf{c_3}$)] $c_{1122}=c_{3311}=c_{2233}=c_{1221}=c_{3113}=c_{2332}=a,c_{2131}=c_{3212}=
                       c_{1323}=6t,b_{1 1 1 1} = 2a^2, b_{1 3 1 2} = -36t^2, b_{2 1 1 2} = -a^2,  b_{2 2 1 1} = -a^2, b_{2 2 2 2} = 2a^2, b_{2 3 1 3} = 12ta, b_{2 3 2 1} = -36t^2, b_{3 1 1 3} = -a^2, b_{3 1 2 1} = 12ta, b_{3 2 1 2} = 12ta, b_{3 2 2 3} = -a^2, b_{3 2 3 1} = -36t^2, b_{3 3 1 1} = -a^2, b_{3 3 2 2} = -a^2, b_{3 3 3 3} = 2a^2 $ ($y_7=y_8=y_9=t, y_0=a$).
 \end{enumerate}

\section{Appendix: Proof of Theorem \ref{mainTH}}
\label{ApB}

In this section we systematically use the summation convention over repeated indices. By Remark \ref{remao} we have to prove that identities (\ref{eqbasic'k},\ref{eqbasic'}) with $\ka=\frac1{\sqrt{N}}$  hold.

We first show that (\ref{eqbasic'k}) holds (in fact only under assumption of skew symmetry of $c$, $c_{ijkl}=-c_{klij}$, i.e. relations (\ref{fp4}) are not required). Let
$$
 \{S_{ij},S_{kl}\}= -\pi(S_{ij})S_{kl}=S_{il}S_{ba}c_{kjab}-S_{kj}S_{ba}c_{ilab}+c_{iqkb}S{}_{qj}S_{bl}-c_{pjal}S_{ip}S_{ka}
$$
and
$$
\{S_{ij},S_{kl}\}_{1}=-\pi_1(S_{ij})S_{kl}=\delta_{jk}S_{il}-\delta_{li}S_{kj}
$$
be the bilinear operations on functions corresponding to the bivectors $\pi$ and $\pi_1$ respectively. Then the trilinear operation corresponding to the Schouten bracket $[\pi,\pi_1]$ is given by
$$
\sum_{c.p.}\left\{ \{S_{ij},S_{kl}\},S_{mn}\right\} _{1}+\left\{ \{S_{ij},S_{kl}\}_{1},S_{mn}\right\},
$$
where the sum is taken over the cyclic permutations of the pairs of indices $(i,j), (k,l)$ and $(m,n)$. We have
\begin{align*}
  & \left\{ \{S_{ij},S_{kl}\},S_{mn}\right\} _{1}+\left\{ \{S_{ij},S_{kl}\}_{1},S_{mn}\right\} =\\
 & \left\{ S_{il}S_{ba}c_{kjab}-S_{kj}S_{ba}c_{ilab}+c_{iqkb}S{}_{qj}S_{bl}-c_{pjal}S_{ip}S_{ka},S_{mn}\right\} _{1}+\\
 & \delta_{jk}\left\{ S_{il},S_{mn}\right\} -\delta_{li}\left\{ S_{kj},S_{mn}\right\} =\\
 & S_{il}c_{kjab}\left\{ S_{ba},S_{mn}\right\} _{1}+S_{ba}c_{kjab}\left\{ S_{il},S_{mn}\right\} _{1}-S_{kj}c_{ilab}\left\{ S_{ba},S_{mn}\right\} _{1}-S_{ba}c_{ilab}\left\{ S_{kj},S_{mn}\right\} _{1}+\\
 & +c_{iqkb}S{}_{qj}\left\{ S_{bl},S_{mn}\right\} _{1}+c_{iqkb}S_{bl}\left\{ S_{qj},S_{mn}\right\} _{1}-c_{pjal}S_{ip}\left\{ S_{ka},S_{mn}\right\} _{1}-c_{pjal}S_{ka}\left\{ S_{ip},S_{mn}\right\} _{1}+\\
 & \delta_{jk}(S_{in}S_{ba}c_{mlab}-S_{ml}S_{ba}c_{inab}+c_{iqmb}S{}_{ql}S_{bn}-c_{plan}S_{ip}S_{ma})+\\
 & -\delta_{li}(S_{kn}S_{ba}c_{mjab}-S_{mj}S_{ba}c_{knab}+c_{kqmb}S{}_{qj}S_{bn}-c_{pjan}S_{kp}S_{ma})=\\
 & \stackrel{1.}{S_{il}c_{kjmb}S_{bn}}-\stackrel{1.}{S_{il}c_{kjan}S_{ma}}+
\stackrel{2.}{S_{ba}c_{kjab}\delta_{lm}S_{in}}-\stackrel{2.}{S_{ba}c_{kjab}\delta_{ni}S_{ml}}\\
 & -\stackrel{1.}{S_{kj}c_{ilmb}S_{bn}}+\stackrel{1.}{S_{kj}c_{ilan}S_{ma}}
 -\stackrel{2.}{S_{ba}c_{ilab}\delta_{jm}S_{kn}}+\stackrel{2.}{S_{ba}c_{ilab}\delta_{nk}S_{mj}}\\
 & +\stackrel{3.}{c_{iqkb}S{}_{qj}\delta_{lm}S_{bn}}-\stackrel{1.}{c_{iqkn}S{}_{qj}S_{ml}}
 +\stackrel{3.}{c_{iqkb}S_{bl}\delta_{jm}S_{qn}}-\stackrel{1.}{c_{inkb}S_{bl}S_{mj}}\\
 & -\stackrel{1.}{c_{pjml}S_{ip}S_{kn}}+\stackrel{3.}{c_{pjal}S_{ip}\delta_{nk}S_{ma}}
 -\stackrel{1.}{c_{mjal}S_{ka}S_{in}}+\stackrel{3.}{c_{pjal}S_{ka}\delta_{ni}S_{mp}}+\\
& \delta_{jk}(\stackrel{2.}{S_{in}S_{ba}c_{mlab}}-\stackrel{2.}{S_{ml}S_{ba}c_{inab}}
+\stackrel{3.}{c_{iqmb}S{}_{ql}S_{bn}}-\stackrel{3.}{c_{plan}S_{ip}S_{ma}})+\\
 & -\delta_{li}(\stackrel{2.}{S_{kn}S_{ba}c_{mjab}}-\stackrel{2.}{S_{mj}S_{ba}c_{knab}}
 +\stackrel{3.}{c_{kqmb}S{}_{qj}S_{bn}}-\stackrel{3.}{c_{pjan}S_{kp}S_{ma}}).
 \end{align*}
 Below we divide the last terms into three groups and sum up over the cyclic permutations of the pairs of indices $(i,j), (k,l)$ and $(m,n)$ (the letters over particular terms indicate the pairs of terms which pairwise cancel out).

\noindent Group 1.:\begin{align*}
 & \stackrel{x}{S_{il}c_{kjmb}S_{bn}}-\stackrel{p}{S_{il}c_{kjan}S_{ma}}
 -\stackrel{s}{S_{kj}c_{ilmb}S_{bn}}+\stackrel{t}{S_{kj}c_{ilan}S_{ma}}\\
 &+ \stackrel{u}{c_{kniq}S{}_{qj}S_{ml}}-\stackrel{q}{c_{inkb}S_{bl}S_{mj}}
 +\stackrel{r}{c_{mlpj}S_{ip}S_{kn}}-\stackrel{o}{c_{mjal}S_{ka}S_{in}}\\
 & \\
 & +\stackrel{z}{S_{kn}c_{mlib}S_{bj}}-\stackrel{r}{S_{kn}c_{mlaj}S_{ia}}
 -\stackrel{u}{S_{ml}c_{knib}S_{bj}}+\stackrel{y}{S_{ml}c_{knaj}S_{ia}}\\
 & +\stackrel{w}{c_{mjkq}S{}_{ql}S_{in}}-\stackrel{x}{c_{kjmb}S_{bn}S_{il}}
 +\stackrel{v}{c_{inpl}S_{kp}S_{mj}}-\stackrel{t}{c_{ilan}S_{ma}S_{kj}}\\
\\ & +\stackrel{q}{S_{mj}c_{inkb}S_{bl}}-\stackrel{v}{S_{mj}c_{inal}S_{ka}}
-\stackrel{w}{S_{in}c_{mjkb}S_{bl}}+\stackrel{o}{S_{in}c_{mjal}S_{ka}}\\
 & +\stackrel{s}{c_{ilmq}S{}_{qn}S_{kj}}-\stackrel{z}{c_{mlib}S_{bj}S_{kn}}
 +\stackrel{p}{c_{kjpn}S_{mp}S_{il}}-\stackrel{y}{c_{knaj}S_{ia}S_{ml}}=0.
 \end{align*}
\noindent Group 2.:
\begin{align*}
  & \stackrel{y}{\delta_{jk}S_{in}S_{ba}c_{mlab}}-\stackrel{z}{\delta_{jk}S_{ml}S_{ba}c_{inab}}
  -\stackrel{u}{\delta_{li}S_{kn}S_{ba}c_{mjab}}+\stackrel{v}{\delta_{li}S_{mj}S_{ba}c_{knab}}\\
 & +\stackrel{r}{S_{ba}c_{kjab}\delta_{lm}S_{in}}-\stackrel{s}{S_{ba}c_{kjab}\delta_{ni}S_{ml}}
 -\stackrel{t}{S_{ba}c_{ilab}\delta_{jm}S_{kn}}+\stackrel{p}{S_{ba}c_{ilab}\delta_{nk}S_{mj}}\\
 &\\
 & +\stackrel{q}{\delta_{lm}S_{kj}S_{ba}c_{inab}}-\stackrel{r}{\delta_{lm}S_{in}S_{ba}c_{kjab}}
 -\stackrel{p}{\delta_{nk}S_{mj}S_{ba}c_{ilab}}+\stackrel{o}{\delta_{nk}S_{il}S_{ba}c_{mjab}}\\
 & +\stackrel{x}{S_{ba}c_{mlab}\delta_{ni}S_{kj}}-\stackrel{y}{S_{ba}c_{mlab}\delta_{jk}S_{in}}
 -{v}{S_{ba}c_{knab}\delta_{li}S_{mj}}+\stackrel{n}{S_{ba}c_{knab}\delta_{jm}S_{il}}\\
\\ & +\stackrel{s}{\delta_{ni}S_{ml}S_{ba}c_{kjab}}-\stackrel{x}{\delta_{ni}S_{kj}S_{ba}c_{mlab}}
-\stackrel{n}{\delta_{jm}S_{il}S_{ba}c_{knab}}+\stackrel{t}{\delta_{jm}S_{kn}S_{ba}c_{ilab}}\\
 & +\stackrel{z}{S_{ba}c_{inab}\delta_{jk}S_{ml}}-\stackrel{q}{S_{ba}c_{inab}\delta_{lm}S_{kj}}
 -\stackrel{o}{S_{ba}c_{mjab}\delta_{nk}S_{il}}+\stackrel{u}{S_{ba}c_{mjab}\delta_{li}S_{kn}}=0.\\
 \end{align*}
\noindent Group 3.:
 \begin{align*}
  & +\stackrel{x}{\delta_{jk}c_{iqmb}S{}_{ql}S_{bn}}-\stackrel{y}{\delta_{jk}c_{plan}S_{ip}S_{ma}}
  -\stackrel{z}{\delta_{li}c_{kqmb}S{}_{qj}S_{bn}}+\stackrel{u}{\delta_{li}c_{pjan}S_{kp}S_{ma}}\\
 & +\stackrel{v}{c_{iqkb}S{}_{qj}\delta_{lm}S_{bn}}+\stackrel{m}{c_{iqkb}S_{bl}\delta_{jm}S_{qn}}
 +\stackrel{n}{c_{pjal}S_{ip}\delta_{nk}S_{ma}}+\stackrel{o}{c_{pjal}S_{ka}\delta_{ni}S_{mp}}\\
 & \\
 & +\stackrel{v}{\delta_{lm}c_{kqib}S{}_{qn}S_{bj}}-\stackrel{p}{\delta_{lm}c_{pnaj}S_{kp}S_{ia}}
 -\stackrel{q}{\delta_{nk}c_{mqib}S{}_{ql}S_{bj}}+\stackrel{n}{\delta_{nk}c_{plaj}S_{mp}S_{ia}}\\
 & +\stackrel{r}{c_{kqmb}S{}_{ql}\delta_{ni}S_{bj}}+\stackrel{z}{c_{kqmb}S_{bn}\delta_{li}S_{qj}}
 +\stackrel{s}{c_{plan}S_{kp}\delta_{jm}S_{ia}}+\stackrel{y}{c_{plan}S_{ma}\delta_{jk}S_{ip}}\\
\\ & +\stackrel{r}{\delta_{ni}c_{mqkb}S{}_{qj}S_{bl}}-\stackrel{o}{\delta_{ni}c_{pjal}S_{mp}S_{ka}}
-\stackrel{m}{\delta_{jm}c_{iqkb}S{}_{qn}S_{bl}}+\stackrel{s}{\delta_{jm}c_{pnal}S_{ip}S_{ka}}\\
 & +\stackrel{x}{c_{mqib}S{}_{qn}\delta_{jk}S_{bl}}+\stackrel{q}{c_{mqib}S_{bj}\delta_{nk}S_{ql}}
 +\stackrel{u}{c_{pnaj}S_{mp}\delta_{li}S_{ka}}+\stackrel{p}{c_{pnaj}S_{ia}\delta_{lm}S_{kp}}=0.\\
\end{align*}
Now we will show  (\ref{eqbasic'}) under the assumption that (\ref{fp4}) holds. To this end we first calculate the trilinear operation on functions
$$
\sum_{c.p.}\left\{ \{S_{ij},S_{kl}\},S_{mn}\right\}
$$
corresponding to the Schouten bracket $[\pi,\pi]$ of $\pi$ with itself. We have
\begin{align*}
  & \{\{S_{ij},S_{kl}\},S_{mn}\}=\{S_{il}S_{l'k'}c_{kjk'l'}-S_{kj}S_{l'k'}c_{ilk'l'}
  +c_{ij'kl'}S{}_{j'j}S_{l'l}-c_{i'jk'l}S_{ii'}S_{kk'},S_{mn}\}=\\
 & S_{il}c_{kjk'l'}\{S_{l'k'},S_{mn}\}+S_{l'k'}c_{kjk'l'}\{S_{il},S_{mn}\}
 -S_{kj}c_{ilk'l'}\{S_{l'k'},S_{mn}\}-S_{l'k'}c_{ilk'l'}\{S_{kj},S_{mn}\}+\\
 & c_{ij'kl'}S{}_{j'j}\{S_{l'l},S_{mn}\}+c_{ij'kl'}S{}_{l'l}\{S_{j'j},S_{mn}\}
 -c_{i'jk'l}S_{ii'}\{S_{kk'},S_{mn}\}-c_{i'jk'l}S_{kk'}\{S_{ii'},S_{mn}\}=\\
 & S_{il}c_{kjk'l'}(\stackrel{5.}{S_{l'n}S_{ba}c_{mk'ab}}-\stackrel{5.}{S_{mk'}S_{ba}c_{l'nab}}
 +\stackrel{6.}{c_{l'qmb}S{}_{qk'}S_{bn}}-\stackrel{6.}{c_{pk'an}S_{l'p}S_{ma}})\\
 & S_{l'k'}c_{kjk'l'}(\stackrel{1.}{S_{in}S_{ba}c_{mlab}}-\stackrel{1.}{S_{ml}S_{ba}c_{inab}}
 +\stackrel{2.}{c_{iqmb}S{}_{ql}S_{bn}}
 -\stackrel{2.}{c_{plan}S_{ip}S_{ma}})\\
 & -S_{kj}c_{ilk'l'}(\stackrel{5.}{S_{l'n}S_{ba}c_{mk'ab}}-\stackrel{5.}{S_{mk'}S_{ba}c_{l'nab}}
 +\stackrel{6.}{c_{l'qmb}S{}_{qk'}S_{bn}} -\stackrel{6.}{c_{pk'an}S_{l'p}S_{ma}})\\
 & -S_{l'k'}c_{ilk'l'}(\stackrel{1.}{S_{kn}S_{ba}c_{mjab}}-\stackrel{1.}{S_{mj}S_{ba}c_{knab}}
 +\stackrel{2.}{c_{kqmb}S{}_{qj}S_{bn}}
 -\stackrel{2.}{c_{pjan}S_{kp}S_{ma}})\\
 & +c_{ij'kl'}S{}_{j'j}(\stackrel{2.}{S_{l'n}S_{ba}c_{mlab}}-\stackrel{7.}{S_{ml}S_{ba}c_{l'nab}}
 +\stackrel{4.}{c_{l'qmb}S{}_{ql}S_{bn}} -\stackrel{3.}{c_{plan}S_{l'p}S_{ma}})\\
 & +c_{ij'kl'}S{}_{l'l}(\stackrel{2.}{S_{j'n}S_{ba}c_{mjab}}-\stackrel{7.}{S_{mj}S_{ba}c_{j'nab}}+
 \stackrel{4.}{c_{j'qmb}S{}_{qj}S_{bn}} -\stackrel{3.}{c_{pjan}S_{j'p}S_{ma}})\\
 & -c_{i'jk'l}S_{ii'}(\stackrel{7.}{S_{kn}S_{ba}c_{mk'ab}}-\stackrel{2.}{S_{mk'}S_{ba}c_{knab}}
 +\stackrel{3.}{c_{kqmb}S{}_{qk'}S_{bn}}-\stackrel{4.}{c_{pk'an}S_{kp}S_{ma}})\\
 & -c_{i'jk'l}S_{kk'}(\stackrel{7.}{S_{in}S_{ba}c_{mi'ab}}-\stackrel{2.}{S_{mi'}S_{ba}c_{inab}}
 +\stackrel{3.}{c_{iqmb}S{}_{qi'}S_{bn}}-\stackrel{4.}{c_{pi'an}S_{ip}S_{ma}}).\\
 \end{align*}
As previuosly, we will divide the terms above into groups and sum up over the cyclic permutations of the pairs of indices $(i,j), (k,l)$ and $(m,n)$. Groups 1.--3. sum up to zero (with no use of (\ref{fp4})):

\noindent Group 1.:\begin{align*}
&
\stackrel{y}{S_{in}c_{kjk'l'}S_{l'k'}c_{mlab}S_{ba}}-\stackrel{z}{S_{ml}c_{kjk'l'}S_{l'k'}c_{inab}S_{ba}}
-\stackrel{u}{S_{kn}c_{ilk'l'}S_{l'k'}c_{mjab}S_{ba}}+\stackrel{v}{S_{mj}c_{ilk'l'}S_{l'k'}c_{knab}S_{ba}}\\
 & +\stackrel{x}{S_{kj}c_{mlk'l'}S_{l'k'}c_{inab}S_{ba}}-\stackrel{y}{S_{in}c_{mlk'l'}S_{l'k'}c_{kjab}S_{ba}}
 -\stackrel{v}{S_{mj}c_{knk'l'}S_{l'k'}c_{ilab}S_{ba}}+\stackrel{r}{S_{il}c_{knk'l'}S_{l'k'}c_{mjab}S_{ba}}\\
 & +\stackrel{z}{S_{ml}c_{ink'l'}S_{l'k'}c_{kjab}S_{ba}}-\stackrel{x}{S_{kj}c_{ink'l'}S_{l'k'}c_{mlab}S_{ba}}
 -\stackrel{r}{S_{il}c_{mjk'l'}S_{l'k'}c_{knab}S_{ba}}+\stackrel{u}{S_{kn}c_{mjk'l'}S_{l'k'}c_{ilab}S_{ba}}=0.\\
 \end{align*}

\noindent Group 2.:\begin{align*}
& \stackrel{x}{c_{kjrs}S_{sr}c_{iqmb}S_{ql}S_{bn}}-\stackrel{y}{c_{kjrs}S_{sr}c_{plan}S_{ip}S_{ma}}
-\stackrel{z}{c_{ilrs}S_{sr}c_{kqmb}S_{qj}S_{bn}}+\stackrel{u}{c_{ilrs}S_{sr}c_{pjan}S_{kp}S_{ma}}\\
 & + \stackrel{v}{c_{mlab}S_{ba}c_{iqks}S_{qj}S_{sn}}+\stackrel{m}{c_{mjab}S_{ba}c_{iqks}S_{qn}S{}_{sl}}
 +\stackrel{n}{c_{knab}S_{ba}c_{pjrl}S_{ip}S_{mr}}+\stackrel{o}{c_{inab}S_{ba}c_{pjrl}S_{mp}S_{kr}}\\
\\ & +\stackrel{v}{c_{mlrs}S_{sr}c_{kqib}S_{qn}S_{bj}}-\stackrel{p}{c_{mlrs}S_{sr}c_{pnaj}S_{kp}S_{ia}}
-\stackrel{q}{c_{knrs}S_{sr}c_{mqib}S_{ql}S_{bj}}+\stackrel{n}{c_{knrs}S_{sr}c_{plaj}S_{mp}S_{ia}}\\
 & +\stackrel{r}{c_{inab}S_{ba}c_{kqms}S_{ql}S_{sj}}+\stackrel{z}{c_{ilab}S_{ba}c_{kqms}S_{qj}S{}_{sn}}
 +\stackrel{s}{c_{mjab}S_{ba}c_{plrn}S_{kp}S_{ir}}+\stackrel{y}{c_{kjab}S_{ba}c_{plrn}S_{ip}S_{mr}}\\
\\ & +\stackrel{r}{c_{inrs}S_{sr}c_{mqkb}S_{qj}S_{bl}}-\stackrel{o}{c_{inrs}S_{sr}c_{pjal}S_{mp}S_{ka}}
-\stackrel{m}{c_{mjrs}S_{sr}c_{iqkb}S_{qn}S_{bl}}+\stackrel{s}{c_{mjrs}S_{sr}c_{pnal}S_{ip}S_{ka}}\\
 &  +\stackrel{x}{c_{kjab}S_{ba}c_{mqis}S_{qn}S_{sl}}+\stackrel{q}{c_{knab}S_{ba}c_{mqis}S_{ql}S{}_{sj}}
 +\stackrel{u}{c_{ilab}S_{ba}c_{pnrj}S_{mp}S_{kr}}+\stackrel{p}{c_{mlab}S_{ba}c_{pnrj}S_{kp}S_{ir}}=0.\\
 \end{align*}

\noindent Group 3.:\begin{align*}
  & -\stackrel{x}{S_{qj}c_{iqks}S_{sp}c_{plan}S_{ma}}+\stackrel{y}{S_{sl}c_{ksiq}S_{qp}c_{pjan}S_{ma}}
  -\stackrel{z}{S_{bn}c_{mbkq}S_{qr}c_{rlpj}S_{ip}}+\stackrel{u}{S_{bn}c_{mbiq}S_{qp}c_{pjrl}S_{kr}}\\
 & -\stackrel{v}{S_{ql}c_{kqms}S_{sp}c_{pnaj}S_{ia}}+\stackrel{z}{S_{sn}c_{mskq}S_{qp}c_{plaj}S_{ia}}
 -\stackrel{w}{S_{bj}c_{ibmq}S_{qr}c_{rnpl}S_{kp}}+\stackrel{x}{S_{bj}c_{ibkq}S_{qp}c_{plrn}S_{mr}}\\
 & -\stackrel{u}{S_{qn}c_{mqis}S_{sp}c_{pjal}S_{ka}}+\stackrel{w}{S_{sj}c_{ismq}S_{qp}c_{pnal}S_{ka}}
 -\stackrel{y}{S_{bl}c_{kbiq}S_{qr}c_{rjpn}S_{mp}}+\stackrel{v}{S_{bl}c_{kbmq}S_{qp}c_{pnrj}S_{ir}}=0.\\
 \end{align*}
We rewrite equations (\ref{fp4}) in the form (recall that $b_{ijkl}=b_{klij}$)
\begin{align}
\label{CB1}c_{klir}c_{rjmn}+c_{ijmr}c_{rnkl}+c_{mnkr}c_{rlij}=\\
\label{CB2}\delta_{kn}b_{mlij}+\delta_{mj}b_{inkl}+\delta_{il}b_{kjmn}\\
\label{CB3}+\frac1{{N}}(\de_{ij}c_{klrs}c_{srmn}+
\de_{mn}c_{ijrs}c_{srkl}+\de_{kl}c_{mnrs}c_{srij})\\
\label{CB4}-\frac{2}{{N}}(\de_{mn}b_{ijkl}+\de_{ij}b_{klmn}+\de_{kl}b_{mnij})
\end{align}
and prepare the terms of groups 4.--7. (summed up over  the cyclic permutations of the pairs of indices $(i,j), (k,l)$ and $(m,n)$) in such a way that  terms of type (\ref{CB1}) appear explicitly.

\noindent Group 4.:\begin{align}
  & S_{i'j}c_{ii'kr}c_{rk'mm'}S{}_{k'l}S_{m'n}-S_{k'l}c_{kk'ip}c_{pi'mm'}S{}_{i'j}S_{m'n}
  +S_{ij'}c_{n'nl's}c_{slj'j}S_{kl'}S_{mn'}-S_{kl'}c_{n'nj'q}c_{qjl'l}S_{ij'}S_{mn'}\nonumber\\
 & S_{k'l}c_{kk'mr}c_{rm'ii'}S{}_{m'n}S_{i'j}-S_{m'n}c_{mm'kp}c_{pk'ii'}S{}_{k'l}S_{i'j}
 +S_{kl'}c_{j'jm's}c_{snl'l}S_{mm'}S_{ij'}-S_{mn'}c_{j'jl'q}c_{qln'n}S_{kl'}S_{ij'}\nonumber\\
 & S_{m'n}c_{mm'ir}c_{ri'kk'}S{}_{i'j}S_{k'l}-S_{i'j}c_{ii'mp}c_{pm'kk'}S{}_{m'n}S_{k'l}
 +S_{mn'}c_{l'lj's}c_{sjn'n}S_{ij'}S_{kl'}-S_{ij'}c_{l'ln'q}c_{qnj'j}S_{mn'}S_{kl'}=\nonumber\\
\nonumber \\ &
 S_{i'j}S{}_{k'l}S_{m'n}(c_{ii'kr}c_{rk'mm'}+c_{kk'mr}c_{rm'ii'}+c_{mm'ir}c_{ri'kk'}
 -c_{kk'ip}c_{pi'mm'}-c_{mm'kp}c_{pk'ii'}-c_{ii'mp}c_{pm'kk'})+\nonumber\\ \label{gr4}
 & S_{ij'}S_{kl'}S_{mn'}(c_{n'nl's}c_{slj'j}+c_{j'jm's}c_{snl'l}+c_{l'lj's}c_{sjn'n}
 -c_{n'nj'q}c_{qjl'l}-c_{j'jl'q}c_{qln'n}-c_{l'ln'q}c_{qnj'j}).
 \end{align}

\noindent Group 5.:\begin{align}
  & S_{il}c_{k'k''mr}c_{rm'kj}S_{k''k'}S_{m'n}-S_{il}c_{kjn'r}c_{rnl''l'}S_{l'l''}S_{mn'}
  -S_{kj}c_{k'k''mr}c_{rm'il}S_{k''k'}S_{m'n}+S_{kj}c_{iln'r}c_{rnl''l'}S_{l'l''}S_{mn'}\nonumber\\
 & S_{kn}c_{m'm''ir}c_{ri'ml}S_{m''m'}S_{i'j}-S_{kn}c_{mlj'r}c_{rjn''n'}S_{n'n''}S_{ij'}
 -S_{ml}c_{m'm''ir}c_{ri'kn}S_{m''m'}S_{i'j}+S_{ml}c_{knj'r}c_{rjn''n'}S_{n'n''}S_{ij'}\nonumber\\
 & S_{mj}c_{i'i''kr}c_{rk'in}S_{i''i'}S_{k'l}-S_{mj}c_{inl'r}c_{rlj''j'}S_{j'j''}S_{kl'}
 -S_{in}c_{i'i''kr}c_{rk'mj}S_{i''i'}S_{k'l}+S_{in}c_{mjl'r}c_{rlj''j'}S_{j'j''}S_{kl'}=\nonumber\\
 \nonumber \\
 & S_{il}S_{k''k'}S_{m'n}(c_{k'k''mr}c_{rm'kj}+c_{kjk'r}c_{rk''mm'}+c_{mm'kr}c_{rjk'k''})\nonumber\\
 &-S_{il}S_{l'l''}S_{mn'}(c_{kjn'r}c_{rnl''l'}+c_{n'nl''r}c_{rl'kj}+c_{l''l'kr}c_{rjn'n})\nonumber\\
 & -S_{kj}S_{k''k'}S_{m'n}(c_{k'k''mr}c_{rm'il}+c_{ilk'r}c_{rk''mm'}+c_{mm'ir}c_{rlk'k''})\nonumber\\
 \label{gr5}
 &+S_{kj}S_{l'l''}S_{mn'}(c_{iln'r}c_{rnl''l'}+c_{n'nl''r}c_{rl'il}+c_{l''l'ir}c_{rln'n}).
  \end{align}

\noindent Group 6.:\begin{align}
  & S_{il}c_{kjk'r}c_{rk''mm'}S{}_{k''k'}S_{m'n}-S_{il}c_{n'nl''r}c_{rl'kj}S_{l'l''}S_{mn'}
  -S_{kj}c_{ilk'r}c_{rk''mm'}S{}_{k''k'}S_{m'n}+S_{kj}c_{n'nl''r}c_{rl'il}S_{l'l''}S_{mn'}\nonumber\\
 & S_{kn}c_{mlm'r}c_{rm''ii'}S{}_{m''m'}S_{i'j}-S_{kn}c_{j'jn''r}c_{rn'ml}S_{n'n''}S_{ij'}
 -S_{ml}c_{knm'r}c_{rm''ii'}S{}_{m''m'}S_{i'j}+S_{ml}c_{j'jn''r}c_{rn'kn}S_{n'n''}S_{ij'}\nonumber\\
 & S_{mj}c_{ini'r}c_{ri''kk'}S{}_{i''i'}S_{k'l}-S_{mj}c_{l'lj''r}c_{rj'in}S_{j'j''}S_{kl'}
 -S_{in}c_{mji'r}c_{ri''kk'}S{}_{i''i'}S_{k'l}+S_{in}c_{l'lj''r}c_{rj'mj}S_{j'j''}S_{kl'}\nonumber\\
 \nonumber \\
 & S_{kn}S_{m''m'}S_{i'j}(c_{m'm''ir}c_{ri'ml}+c_{mlm'r}c_{rm''ii'}+c_{ii'mr}c_{rlm'm''})\nonumber\\
 &-S_{kn}S_{n'n''}S_{ij'}(c_{mlj'r}c_{rjn''n'}+c_{j'jn''r}c_{rn'ml}+c_{n''n'mr}c_{rlj'j})\nonumber\\
 & -S_{ml}S_{m''m'}S_{i'j}(c_{m'm''ir}c_{ri'kn}+c_{knm'r}c_{rm''ii'}+c_{ii'kr}c_{rnm'm''})\nonumber\\
 \label{gr6}
 &+S_{ml}S_{n'n''}S_{ij'}(c_{knj'r}c_{rjn''n'}+c_{j'jn''r}c_{rn'kn}+c_{n''n'kr}c_{rnj'j}).
 \end{align}

\noindent Group 7.:\begin{align}
& -S_{ml}c_{ii'kr}c_{rnm'm''}S_{m''m'}S_{i'j}+S_{mj}c_{kk'ir}c_{rni'i''}S_{i''i'}S_{k'l}
-S_{kn}c_{n''n'mr}c_{rlj'j}S_{n'n''}S_{ij'}+S_{in}c_{j''j'mr}c_{rjl'l}S_{j'j''}S_{kl'}\nonumber\\
 & -S_{in}c_{kk'mr}c_{rji'i''}S_{i''i'}S_{k'l}+S_{il}c_{mm'kr}c_{rjk'k''}S_{k''k'}S_{m'n}
 -S_{mj}c_{j''j'ir}c_{rnl'l}S_{j'j''}S_{kl'}+S_{kj}c_{l''l'ir}c_{rln'n}S_{l'l''}S_{mn'}\nonumber\\
 & -S_{kj}c_{mm'ir}c_{rlk'k''}S_{k''k'}S_{m'n}+S_{kn}c_{ii'mr}c_{rlm'm''}S_{m''m'}S_{i'j}
 -S_{il}c_{l''l'kr}c_{rjn'n}S_{l'l''}S_{mn'}+S_{ml}c_{n''n'kr}c_{rnj'j}S_{n'n''}S_{ij'}\nonumber\\
 \nonumber \\
 & S_{mj}S_{i''i'}S_{k'l}(c_{i'i''kr}c_{rk'in}+c_{ini'r}c_{ri''kk'}+c_{kk'ir}c_{rni'i''})\nonumber\\
 &-S_{mj}S_{j'j''}S_{kl'}(c_{inl'r}c_{rlj''j'}+c_{l'lj''r}c_{rj'in}+c_{j''j'ir}c_{rnl'l})\nonumber \\
 & -S_{in}S{}_{i''i'}S_{k'l}(c_{i'i''kr}c_{rk'mj}+c_{mji'r}c_{ri''kk'}+c_{kk'mr}c_{rji'i''})\nonumber \\\label{gr7}
 &+S_{in}S_{j'j''}S_{kl'}(c_{mjl'r}c_{rlj''j'}+c_{l'lj''r}c_{rj'mj}+c_{j''j'mr}c_{rjl'l}).
 \end{align}
Now, in the sum
\begin{align}
&
 S_{i'j}S{}_{k'l}S_{m'n}(c_{ii'kr}c_{rk'mm'}+c_{kk'mr}c_{rm'ii'}+c_{mm'ir}c_{ri'kk'}
 -c_{kk'ip}c_{pi'mm'}-c_{mm'kp}c_{pk'ii'}-c_{ii'mp}c_{pm'kk'})\nonumber\\
 & +S_{ij'}S_{kl'}S_{mn'}(c_{n'nl's}c_{slj'j}+c_{j'jn's}c_{snl'l}+c_{l'lj's}c_{sjn'n}
 -c_{n'nj'q}c_{qjl'l}-c_{j'jl'q}c_{qln'n}-c_{l'ln'q}c_{qnj'j})\nonumber\\
 & +S_{il}S_{k''k'}S_{m'n}(c_{k'k''mr}c_{rm'kj}+c_{kjk'r}c_{rk''mm'}+c_{mm'kr}c_{rjk'k''})\nonumber\\
 &-S_{il}S_{l'l''}S_{mn'}(c_{kjn'r}c_{rnl''l'}+c_{n'nl''r}c_{rl'kj}+c_{l''l'kr}c_{rjn'n})\nonumber\\
 & -S_{kj}S_{k''k'}S_{m'n}(c_{k'k''mr}c_{rm'il}+c_{ilk'r}c_{rk''mm'}+c_{mm'ir}c_{rlk'k''})\nonumber\\
 &+S_{kj}S_{l'l''}S_{mn'}(c_{iln'r}c_{rnl''l'}+c_{n'nl''r}c_{rl'il}+c_{l''l'ir}c_{rln'n})\nonumber\\
& +S_{kn}S_{m''m'}S_{i'j}(c_{m'm''ir}c_{ri'ml}+c_{mlm'r}c_{rm''ii'}+c_{ii'mr}c_{rlm'm''})\nonumber\\
 &-S_{kn}S_{n'n''}S_{ij'}(c_{mlj'r}c_{rjn''n'}+c_{j'jn''r}c_{rn'ml}+c_{n''n'mr}c_{rlj'j})\nonumber\\
 & -S_{ml}S_{m''m'}S_{i'j}(c_{m'm''ir}c_{ri'kn}+c_{knm'r}c_{rm''ii'}+c_{ii'kr}c_{rnm'm''})\nonumber\\
  &+S_{ml}S_{n'n''}S_{ij'}(c_{knj'r}c_{rjn''n'}+c_{j'jn''r}c_{rn'kn}+c_{n''n'kr}c_{rnj'j})
 \nonumber \\
 &+ S_{mj}S_{i''i'}S_{k'l}(c_{i'i''kr}c_{rk'in}+c_{ini'r}c_{ri''kk'}+c_{kk'ir}c_{rni'i''})\nonumber\\
 &-S_{mj}S_{j'j''}S_{kl'}(c_{inl'r}c_{rlj''j'}+c_{l'lj''r}c_{rj'in}+c_{j''j'ir}c_{rnl'l})\nonumber \\
 & -S_{in}S{}_{i''i'}S_{k'l}(c_{i'i''kr}c_{rk'mj}+c_{mji'r}c_{ri''kk'}+c_{kk'mr}c_{rji'i''})\nonumber \\\label{gr4-7}
 &+S_{in}S_{j'j''}S_{kl'}(c_{mjl'r}c_{rlj''j'}+c_{l'lj''r}c_{rj'mj}+c_{j''j'mr}c_{rjl'l}).
 \end{align}
of expressions (\ref{gr4}--\ref{gr7}) the terms of type (\ref{CB1}) are substituted by the corresponding terms of type (\ref{CB2}), the result cancels out:
\begin{align*}
 &S_{i'j}S{}_{k'l}S_{m'n}(b_{ii'km'}\delta_{mk'}+b_{mm'ik'}\delta_{ki'}+b_{kk'mi'}\delta_{im'}
 -b_{kk'im'}\delta_{mi'}-b_{ii'mk'}\delta_{km'}-b_{mm'ki'}\delta_{ik'})\\
 & +S_{ij'}S_{kl'}S_{mn'}(b_{n'nl'j}\delta_{j'l}+b_{j'jn'l}\delta_{l'n}+b_{l'lj'n}\delta_{n'j}
 -b_{j'jl'n}\delta_{n'l}-b_{n'nj'l}\delta_{l'j}-b_{l'ln'j}\delta_{j'n})\\
& +S_{il}S_{k''k'}S_{m'n}(b_{k'k''mj}\delta_{km'}+b_{kjk'm'}\delta_{mk''}+b_{mm'kk''}\delta_{k'j})-
S_{il}S_{l'l''}S_{mn'}(b_{kjn'l'}\delta_{l''n}+b_{n'nl''j}\delta_{kl'}+b_{l''l'kn}\delta_{n'j})\\
& -S_{kj}S_{k''k'}S_{m'n}(b_{k'k''ml}\delta_{im'}+b_{ilk'm'}\delta_{mk''}+b_{mm'ik''}\delta_{k'l})+
S_{kj}S_{l'l''}S_{mn'}(b_{iln'l'}\delta_{l''n}+b_{n'nl''l}\delta_{il'}+b_{l''l'in}\delta_{n'l})\\
& +S_{kn}S_{m''m'}S_{i'j}(b_{m'm''il}\delta_{mi'}+b_{mlm'i'}\delta_{im''}+b_{ii'mm''}\delta_{m'l})
-S_{kn}S_{n'n''}S_{ij'}(b_{mlj'n'}\delta_{n''j}+b_{j'jn''l}\delta_{mn'}+b_{n''n'mj}\delta_{j'l})\\
& -S_{ml}S_{m''m'}S_{i'j}(b_{m'm''in}\delta_{ki'}+b_{knm'i'}\delta_{im''}+b_{ii'km''}\delta_{m'n})
+S_{ml}S_{n'n''}S_{ij'}(b_{knj'n'}\delta_{n''j}+b_{j'jn''n}\delta_{kn'}+b_{n''n'kj}\delta_{j'n})\\
& +S_{mj}S_{i''i'}S_{k'l}(b_{i'i''kn}\delta_{ik'}+b_{ini'k'}\delta_{ki''}+b_{kk'ii''}\delta_{i'n})
-S_{mj}S_{j'j''}S_{kl'}(b_{inl'j'}\delta_{j''l}+b_{l'lj''n}\delta_{ij'}+b_{j''j'il}\delta_{l'n})\\
& -S_{in}S{}_{i''i'}S_{k'l}(b_{i'i''kj}\delta_{mk'}+b_{mji'k'}\delta_{ki''}+b_{kk'mi''}\delta_{i'j})
+S_{in}S_{j'j''}S_{kl'}(b_{mjl'j'}\delta_{j''l}+b_{l'lj''j}\delta_{mj'}+b_{j''j'ml}\delta_{l'j})=\\
  & +\stackrel{f}{S_{ml}b_{ii'km'}S_{i'j}S_{m'n}}+\stackrel{v}{S_{kj}b_{mm'ik'}S_{m'n}S_{k'l}}+
  \stackrel{m}{S_{in}b_{kk'mi'}S{}_{k'l}S_{i'j}}\\
 & -\stackrel{j}{S_{mj}b_{kk'im'}S{}_{k'l}S_{m'n}}-\stackrel{c}{S_{kn}b_{ii'mk'}S_{i'j}S{}_{k'l}}
 -\stackrel{y}{S{}_{il}b_{mm'ki'}S_{m'n}S_{i'j}}\\
 & +\stackrel{z}{S_{il}b_{n'nl'j}S_{mn'}S_{kl'}}+\stackrel{d}{S_{kn}b_{j'jn'l}S_{ij'}S_{mn'}}
 +\stackrel{k}{S_{mj}b_{l'lj'n}S_{kl'}S_{ij'}}\\
 & -\stackrel{g}{S_{ml}b_{j'jl'n}S_{ij'}S_{kl'}}-\stackrel{a}{S_{kj}b_{n'nj'l}S_{mn'}S_{ij'}}
 -\stackrel{n}{S_{in}b_{l'ln'j}S_{kl'}S_{mn'}}
\\ & +\stackrel{h}{S_{il}S_{k''k'}b_{k'k''mj}S_{kn}}+\stackrel{x}{S_{il}b_{kjk'm'}S_{mk'}S_{m'n}}
+\stackrel{y}{S_{il}b_{mm'kk''}S_{m'n}S_{k''j}}\\ &
-\stackrel{x}{S_{il}b_{kjn'l'}S_{mn'}S_{l'n}}-\stackrel{z}{S_{il}b_{n'nl''j}S_{mn'}S_{kl''}}
-\stackrel{o}{S_{il}S_{l'l''}b_{l''l'kn}S_{mj}}\\
 & -\stackrel{p}{S_{kj}S_{k''k'}S_{in}b_{kk''ml}}-\stackrel{u}{S_{kj}b_{ilk'm'}S_{mk'}S_{m'n}}-
 \stackrel{v}{S_{kj}b_{mm'ik''}S_{m'n}S_{k''l}}\\ &
 +\stackrel{u}{S_{kj}b_{iln'l'}S_{mn'}S_{l'n}}+\stackrel{a}{S_{kj}b_{n'nl''l}S_{mn'}S_{il''}}
 +\stackrel{q}{S_{kj}S_{l'l''}S_{ml}b_{l''l'in}}\\
 & +\stackrel{r}{S_{kn}S_{m''m'}S_{mj}b_{m'm''il}}+\stackrel{b}{S_{kn}b_{mlm'i'}S_{im'}S_{i'j}}
 +\stackrel{c}{S_{kn}b_{ii'mm''}S_{i'j}S_{m''l}}\\ &
 -\stackrel{b}{S_{kn}b_{mlj'n'}S_{ij'}S_{n'j}}-\stackrel{d}{S_{kn}b_{j'jn''l}S_{ij'}S_{mn''}}
 -\stackrel{h}{S_{kn}S_{n'n''}S_{il}b_{n''n'mj}}\\
 & -\stackrel{q}{S_{ml}S_{m''m'}S_{kj}b_{m'm''in}}-\stackrel{e}{S_{ml}b_{knm'i'}S_{im'}S_{i'j}}
 -\stackrel{f}{S_{ml}b_{ii'km''}S_{i'j}S_{m''n}}\\ &
 +\stackrel{e}{S_{ml}b_{knj'n'}S_{ij'}S_{n'j}}+\stackrel{g}{S_{ml}b_{j'jn''n}S_{ij'}S_{kn''}}
 +\stackrel{s}{S_{ml}S_{n'n''}S_{in}b_{n''n'kj}}\\
 & +\stackrel{o}{S_{mj}S_{i''i'}S_{il}b_{i'i''kn}}+\stackrel{i}{S_{mj}b_{ini'k'}S_{ki'}S_{k'l}}
 +\stackrel{j}{S_{mj}b_{kk'ii''}S_{k'l}S_{i''n}}\\ &
 -\stackrel{i}{S_{mj}b_{inl'j'}S_{kl'}S_{j'l}}-\stackrel{k}{S_{mj}b_{l'lj''n}S_{kl'}S_{ij''}}
 -\stackrel{r}{S_{mj}S_{j'j''}S_{kn}b_{j''j'il}}\\
 & -\stackrel{s}{S_{in}S{}_{i''i'}S_{ml}b_{i'i''kj}}-\stackrel{l}{S_{in}b_{mji'k'}S_{ki'}S_{k'l}}
 -\stackrel{m}{S_{in}b_{kk'mi''}S_{k'l}S_{i''j}}\\ &
 +\stackrel{l}{S_{in}b_{mjl'j'}S_{kl'}S_{j'l}}+\stackrel{n}{S_{in}b_{l'lj''j}S_{kl'}S_{mj''}}
 +\stackrel{p}{S_{in}S_{j'j''}S_{kj}b_{j''j'ml}}=0.
\end{align*}
Analogously, in expression (\ref{gr4-7}) we now substitute the terms of type (\ref{CB1}) by terms of type (\ref{CB4}). We get
\begin{align*}
 &-\frac{2}{N}(
 S_{i'j}S{}_{k'l}S_{m'n}(\de_{ii'}b_{kk'mm'}+\de_{kk'}b_{mm'ii'}+\de_{mm'}b_{ii'kk'}
 -\de_{kk'}b_{ii'mm'}-\de_{mm'}b_{kk'ii'}-\de_{ii'}b_{mm'kk'})\nonumber\\
 & +S_{ij'}S_{kl'}S_{mn'}(\de_{n'n}b_{l'lj'j}+\de_{j'j}b_{n'nl'l}+\de_{l'l}b_{j'jn'n}
 -\de_{n'n}b_{j'jl'l}-\de_{j'j}b_{l'ln'n}-\de_{l'l}b_{n'nj'j})\nonumber\\
 & +S_{il}S_{k''k'}S_{m'n}(\de_{k'k''}b_{mm'kj}+\de_{kj}b_{k'k''mm'}+\de_{mm'}b_{kjk'k''})\nonumber\\
 &-S_{il}S_{l'l''}S_{mn'}(\de_{kj}b_{n'nl''l'}+\de_{n'n}b_{l''l'kj}+\de_{l''l'}b_{kjn'n})\nonumber\\
 & -S_{kj}S_{k''k'}S_{m'n}(\de_{k'k''}b_{mm'il}+\de_{il}b_{k'k''mm'}+\de_{mm'}b_{ilk'k''})\nonumber\\
 &+S_{kj}S_{l'l''}S_{mn'}(\de_{il}b_{n'nl''l'}+\de_{n'n}b_{l''l'il}+\de_{l''l'}b_{iln'n})\nonumber\\
& +S_{kn}S_{m''m'}S_{i'j}(\de_{m'm''}b_{ii'ml}+\de_{ml}b_{m'm''ii'}+\de_{ii'}b_{mlm'm''})\nonumber\\
 &-S_{kn}S_{n'n''}S_{ij'}(\de_{ml}b_{j'jn''n'}+\de_{j'j}b_{n''n'ml}+\de_{n''n'}b_{mlj'j})\nonumber\\
 & -S_{ml}S_{m''m'}S_{i'j}(\de_{m'm''}b_{ii'kn}+\de_{kn}b_{m'm''ii'}+\de_{ii'}b_{knm'm''})\nonumber\\
  &+S_{ml}S_{n'n''}S_{ij'}(\de_{kn}b_{j'jn''n'}+\de_{j'j}b_{n''n'kn}+\de_{n''n'}b_{knj'j})
 \nonumber \\
 &+ S_{mj}S_{i''i'}S_{k'l}(\de_{i'i''}b_{kk'in}+\de_{in}b_{i'i''kk'}+\de_{kk'}b_{ini'i''})\nonumber\\
 &-S_{mj}S_{j'j''}S_{kl'}(\de_{in}b_{l'lj''j'}+\de_{l'l}b_{j''j'in}+\de_{j''j'}b_{inl'l})\nonumber \\
 & -S_{in}S{}_{i''i'}S_{k'l}(\de_{i'i''}b_{kk'mj}+\de_{mj}b_{i'i''kk'}+\de_{kk'}b_{mji'i''})\nonumber \\
 &+S_{in}S_{j'j''}S_{kl'}(\de_{mj}b_{l'lj''j'}+\de_{l'l}b_{j''j'mj}+\de_{j''j'}b_{mjl'l}))=\nonumber \\
 &-\frac{2}{N}(
 \stackrel{a}{S_{ij}S{}_{k'l}S_{m'n}b_{kk'mm'}}+\stackrel{b}{S_{i'j}S{}_{kl}S_{m'n}b_{mm'ii'}}
 +\stackrel{c}{S_{i'j}S{}_{k'l}S_{mn}b_{ii'kk'}}\nonumber\\
 &-\stackrel{b}{S_{i'j}S{}_{kl}S_{m'n}b_{ii'mm'}}-\stackrel{c}{S_{i'j}S{}_{k'l}S_{mn}b_{kk'ii'}}
 -\stackrel{a}{S_{ij}S{}_{k'l}S_{m'n}b_{mm'kk'}}\nonumber\\
 & +\stackrel{e}{S_{ij'}S_{kl'}S_{mn'}b_{l'lj'j}}+\stackrel{d}{S_{ij}S_{kl'}S_{mn}b_{n'nl'l}}+
 \stackrel{f}{S_{ij'}S_{kl}S_{mn}b_{j'jn'n}}\nonumber\\
 &-\stackrel{e}{S_{ij'}S_{kl'}S_{mn'}b_{j'jl'l}}-\stackrel{d}{S_{ij}S_{kl'}S_{mn}b_{l'ln'n}}
 -\stackrel{f}{S_{ij'}S_{kl}S_{mn}b_{n'nj'j}}\nonumber\\
 & +S_{il}S_{k'k'}S_{m'n}b_{mm'kj}+S_{il}S_{k''k'}S_{m'n}\de_{kj}b_{k'k''mm'}+
 \stackrel{g}{S_{il}S_{k''k'}S_{mn}b_{kjk'k''}}\nonumber\\
 &- S_{il}S_{l'l''}S_{mn'}\de_{kj}b_{n'nl''l'}- \stackrel{g}{S_{il}S_{l'l''}S_{mn}b_{l''l'kj}}
 -S_{il}S_{l'l'}S_{mn'}b_{kjn'n}\nonumber\\
 & -S_{kj}S_{k'k'}S_{m'n}b_{mm'il}-S_{kj}S_{k''k'}S_{m'n}\de_{il}b_{k'k''mm'}
 -\stackrel{i}{S_{kj}S_{k''k'}S_{mn}b_{ilk'k''}}\nonumber\\
 &+S_{kj}S_{l'l''}S_{mn'}\de_{il}b_{n'nl''l'}+\stackrel{i}{S_{kj}S_{l'l''}S_{mn}b_{l''l'il}}
 +S_{kj}S_{l'l'}S_{mn'}b_{iln'n}\nonumber\\
& +S_{kn}S_{m'm'}S_{i'j}b_{ii'ml}+S_{kn}S_{m''m'}S_{i'j}\de_{ml}b_{m'm''ii'}
+\stackrel{j}{S_{kn}S_{m''m'}S_{ij}b_{mlm'm''}}\nonumber\\
 &-S_{kn}S_{n'n''}S_{ij'}\de_{ml}b_{j'jn''n'}-\stackrel{j}{S_{kn}S_{n'n''}S_{ij}b_{n''n'ml}}
 -S_{kn}S_{n'n'}S_{ij'}b_{mlj'j}\nonumber\\
 & -S_{ml}S_{m'm'}S_{i'j}b_{ii'kn}-S_{ml}S_{m''m'}S_{i'j}\de_{kn}b_{m'm''ii'}
 -\stackrel{k}{S_{ml}S_{m''m'}S_{ij}b_{knm'm''}}\nonumber\\
  &+S_{ml}S_{n'n''}S_{ij'}\de_{kn}b_{j'jn''n'}+\stackrel{k}{S_{ml}S_{n'n''}S_{ij}b_{n''n'kn}}
  +S_{ml}S_{n'n'}S_{ij'}b_{knj'j}
 \nonumber \\
 &+ S_{mj}S_{i'i'}S_{k'l}b_{kk'in}+S_{mj}S_{i''i'}S_{k'l}\de_{in}b_{i'i''kk'}
 +\stackrel{l}{S_{mj}S_{i''i'}S_{kl}b_{ini'i''}}\nonumber\\
 &-S_{mj}S_{j'j''}S_{kl'}\de_{in}b_{l'lj''j'}-\stackrel{l}{S_{mj}S_{j'j''}S_{kl}b_{j''j'in}}
 -S_{mj}S_{j'j'}S_{kl'}b_{inl'l}\nonumber \\
 & -S_{in}S{}_{i'i'}S_{k'l}b_{kk'mj}-S_{in}S{}_{i''i'}S_{k'l}\de_{mj}b_{i'i''kk'}
 -\stackrel{m}{S_{in}S{}_{i''i'}S_{kl}b_{mji'i''}}\nonumber \\
 &+S_{in}S_{j'j''}S_{kl'}\de_{mj}b_{l'lj''j'}+\stackrel{m}{S_{in}S_{j'j''}S_{kl}b_{j''j'mj}}
 +S_{in}S_{j'j'}S_{kl'}b_{mjl'l}).
  \end{align*}
As previously, we indicate by letters above the terms those that pairwise cancel out. After the restriction to $sl(N)$ (i.e. after putting $S_{ii}=0$), the following terms remain:
\begin{align}\nonumber
  &-\frac{2}{N}(\stackrel{a}{S_{il}S_{k''k'}S_{m'n}\de_{kj}b_{k'k''mm'}}
 - \stackrel{b}{S_{il}S_{l'l''}S_{mn'}\de_{kj}b_{n'nl''l'}}\\ \nonumber
 & -\stackrel{c}{S_{kj}S_{k''k'}S_{m'n}\de_{il}b_{k'k''mm'}}
  +\stackrel{d}{S_{kj}S_{l'l''}S_{mn'}\de_{il}b_{n'nl''l'}}\\ \nonumber
&+\stackrel{e}{S_{kn}S_{m''m'}S_{i'j}\de_{ml}b_{m'm''ii'}}
 -\stackrel{f}{S_{kn}S_{n'n''}S_{ij'}\de_{ml}b_{j'jn''n'}}\\ \nonumber
 &-\stackrel{g}{S_{ml}S_{m''m'}S_{i'j}\de_{kn}b_{m'm''ii'}}
   +\stackrel{i}{S_{ml}S_{n'n''}S_{ij'}\de_{kn}b_{j'jn''n'}}+ \\ \nonumber
 &+\stackrel{j}{S_{mj}S_{i''i'}S_{k'l}\de_{in}b_{i'i''kk'}}
  -\stackrel{k}{S_{mj}S_{j'j''}S_{kl'}\de_{in}b_{l'lj''j'}}\\ \label{last}
&-\stackrel{l}{S_{in}S{}_{i''i'}S_{k'l}\de_{mj}b_{i'i''kk'}}
  +\stackrel{m}{S_{in}S_{j'j''}S_{kl'}\de_{mj}b_{l'lj''j'}}).
  \end{align}
Finally, we calculate the trilinear operation on functions corresponding to the trivector $\pi_3':=\pi_1\wedge \pi_1(\frac1{N}H')$, $H'=S_{ls}b_{slnm}S_{mn}$, $\pi_1(\frac1{N} H')=\frac1{N}(2b_{ksnm}S_{mn}S_{sl}-2b_{slnm}S_{mn}S_{ks})\frac{\d}{\d S_{kl }}$.
\begin{align} \nonumber
  &(S_{ij},S_{kl},S_{mn})_{\pi_3'}=\langle dS_{ij}\wedge dS_{kl}\wedge dS_{mn}, \pi_1\wedge \pi_1(\frac1{N}h')\rangle=\\ \nonumber &\frac2{N}((\de_{jk}S_{il}-\de_{li}S_{kj})(b_{msn'm'}S_{m'n'}S_{sn}-b_{snn'm'}S_{m'n'}S_{ms})+\\ \nonumber
 & (\de_{lm}S_{kn}-\de_{nk}S_{ml})(b_{isn'm'}S_{m'n'}S_{sj}-b_{sjn'm'}S_{m'n'}S_{is})+\\ \nonumber
 & (\de_{ni}S_{mj}-\de_{jm}S_{in})(b_{ksn'm'}S_{m'n'}S_{sl}-b_{sln'm'}S_{m'n'}S_{ks}))=\\ \nonumber
 &\frac2{N}(\stackrel{a}{\de_{jk}S_{il}b_{msn'm'}S_{m'n'}S_{sn}}-\stackrel{b}{\de_{jk}S_{il}b_{snn'm'}S_{m'n'}S_{ms}}\\ \nonumber
 &-\stackrel{c}{\de_{li}S_{kj}b_{msn'm'}S_{m'n'}S_{sn}}+\stackrel{d}{\de_{li}S_{kj}b_{snn'm'}S_{m'n'}S_{ms}}\\ \nonumber
 &+\stackrel{e}{\de_{lm}S_{kn}b_{isn'm'}S_{m'n'}S_{sj}}-\stackrel{f}{\de_{lm}S_{kn}b_{sjn'm'}S_{m'n'}S_{is}}\\ \nonumber
 &-\stackrel{g}{\de_{nk}S_{ml}b_{isn'm'}S_{m'n'}S_{sj}}+\stackrel{i}{\de_{nk}S_{ml}b_{sjn'm'}S_{m'n'}S_{is}}\\ \nonumber
 &+\stackrel{j}{\de_{ni}S_{mj}b_{ksn'm'}S_{m'n'}S_{sl}}-\stackrel{k}{\de_{ni}S_{mj}b_{sln'm'}S_{m'n'}S_{ks}}\\ \label{last1}
 &-\stackrel{l}{\de_{jm}S_{in}b_{ksn'm'}S_{m'n'}S_{sl}}+\stackrel{m}{\de_{jm}S_{in}b_{sln'm'}S_{m'n'}S_{ks}}).
  \end{align}
We see that expressions (\ref{last}) and (\ref{last1}) are equal up to sign (as before we indicate the pairs of the corresponding terms by letters above these terms). Analogously one proves that, if  in expression (\ref{gr4-7}) one substitutes the terms of type (\ref{CB1}) by terms of type (\ref{CB3}), then up to sign one gets $(S_{ij},S_{kl},S_{mn})_{\pi_3''}$, where
$\pi_3'':=\pi_1\wedge \pi_1(\frac1{N} H'')$, $H''=-\frac1{2}(S_{ls}c_{sln'm'}c_{m'n'mn}S_{mn})$, $\pi_1(\frac1{N} H'')=\frac1{N}(-c_{ksnm}c_{mnn'm'}S_{m'n'}S_{sl}+
c_{slnm}c_{mnn'm'}S_{m'n'}S_{ks})\frac{\d}{\d S_{kl }}$. Taking into account the coefficient $-2$ in formula (\ref{ghj}) we have proven that $(S_{ij},S_{kl},S_{mn})_{\pi}=2(S_{ij},S_{kl},S_{mn})_{\pi_3'+\pi_3''}$, i.e. (\ref{eqbasic'}) is established and by Remark \ref{remao} this finishes the proof. \qed


\end{document}